\documentclass[%
 reprint,
superscriptaddress,
 amsmath,amssymb,
 aps,
prb,
]{revtex4-2}
\usepackage{graphicx}
\usepackage{dcolumn}
\usepackage{bm}
\usepackage{color}
\usepackage{multirow}
\usepackage{placeins}
\usepackage{setspace}
\usepackage{makecell}
\usepackage{titlesec}
\usepackage{tocloft}
\usepackage[bookmarks=false,dvipdfm]{hyperref}
\hypersetup{
  colorlinks=true,
  linkcolor=black,
  urlcolor=black,
  citecolor=black,
}
\begin{document}
\title{Quantum Criticality by Interaction Frustration in a Square-Planar Lattice}

\author{Yi-Qiang Lin}
\email[corresponding author: ]{yiqianglin@zju.edu.cn}
\affiliation{School of Physics, Zhejiang University, Hangzhou 310058, China}

\author{Chang-Chao Liu}
\affiliation{School of Physics, Zhejiang University, Hangzhou 310058, China}

\author{Jia-Xin Li}
\affiliation{School of Physics, Zhejiang University, Hangzhou 310058, China}

\author{Bai-Jiang Lv}
\affiliation{ Key Laboratory of Neutron Physics and Institute of Nuclear Physics and Chemistry, China Academy of Engineering Physics, Mianyang, Sichuan 621999, China}

\author{Kai-Xin Ye}
\affiliation{Center for Correlated Matter, School of Physics, Zhejiang University, Hangzhou 310058, China}

\author{Jia-Wen Zhang}
\affiliation{Center for Correlated Matter, School of Physics, Zhejiang University, Hangzhou 310058, China}

\author{Si-Qi Wu}
\affiliation{Department of Physics, The Hong Kong University of Science and Technology, Clear Water Bay, Hong Kong, China}

\author{Ya-Nan Zhang}
\affiliation{Center for Correlated Matter, School of Physics, Zhejiang University, Hangzhou 310058, China}

\author{Ye Chen}
\affiliation{Center for Correlated Matter, School of Physics, Zhejiang University, Hangzhou 310058, China}

\author{Jia-Yi Lu}
\affiliation{School of Physics, Zhejiang University, Hangzhou 310058, China}

\author{Jing Li}
\affiliation{School of Physics, Zhejiang University, Hangzhou 310058, China}
 
\author{Hua-Xun Li}
\affiliation{School of Physics, Zhejiang University, Hangzhou 310058, China}

\author{Hao Li}
\affiliation{ Key Laboratory of Neutron Physics and Institute of Nuclear Physics and Chemistry, China Academy of Engineering Physics, Mianyang, Sichuan 621999, China}

\author{Yi Liu}
\affiliation{Department of Applied Physics, Key Laboratory of Quantum Precision Measurement of Zhejiang Province, Zhejiang University of Technology, Hangzhou, China}

\author{Cao Wang}
\affiliation{School of Physics and Optoelectronic Engineering, Shandong University of Technology, Zibo, 255000, Shandong, China}

\author{Yun-Lei Sun}
\affiliation{School of Information and Electrical Engineering, Hangzhou City University, Hangzhou, 310015, Zhejiang, China}

\author{Hao Jiang}
\affiliation{School of Physics and Optoelectronics, Xiangtan University, Xiangtan 411105, China}

\author{Hui-Qiu Yuan}
\affiliation{Center for Correlated Matter, School of Physics, Zhejiang University, Hangzhou 310058, China}
\affiliation{Institute for Advanced Study in Physics, Zhejiang University, Hangzhou 310058, China}
\affiliation{Institute of Fundamental and Transdisciplinary Research, Zhejiang University, Hangzhou 310058, China}
\affiliation{State Key Laboratory of Silicon and Advanced Semiconductor Materials, Zhejiang University, Hangzhou 310058, China}

\author{Guang-Han Cao}
\email[corresponding author: ]{ghcao@zju.edu.cn}
\affiliation{School of Physics, Zhejiang University, Hangzhou 310058, China}
\affiliation{Institute of Fundamental and Transdisciplinary Research, Zhejiang University, Hangzhou 310058, China}
\affiliation{State Key Laboratory of Silicon and Advanced Semiconductor Materials, Zhejiang University, Hangzhou 310058, China}
\affiliation{Collaborative Innovation Center of Advanced Microstructures, Nanjing University, Nanjing, 210093, China}


\begin{abstract}
We report experimental and theoretical investigations on ThCr$_2$Ge$_2$C, a metallic compound in which Cr$_2$C planes form a square-planar lattice. Neutron powder diffraction, magnetization, and specific heat measurements reveal no evidence of long-range magnetic order or short-range spin freezing down to 70~mK. Quantum critical behavior was indicated through logarithmic divergences in both the magnetic susceptibility and the specific heat divided by temperature. Resistivity measurements exhibit non-Fermi-liquid behavior, with a Fermi liquid recovered under magnetic fields or high pressures. First-principles calculations identify competing nearest-neighbor ($J_1$) and next-nearest-neighbor ($J_2$) exchange interactions, with $J_2/J_1 \sim -0.5$, pointing to strong magnetic frustration. The interaction frustration is reduced, and magnetically ordered phases are stabilized upon the application of negative or positive pressures. This work offers a rare example of zero-field, ambient pressure quantum criticality mainly driven by interaction frustration in a square lattice.
\end{abstract}

\maketitle
Quantum phase transitions represent a cornerstone of condensed matter physics, where materials exhibit drastic changes in their ground state at zero temperature, induced by variations in non-thermal control parameters, e.g., pressure, magnetic field, or chemical doping~\cite{1999_QPT,2003_QPT,2008-quantumcriticality,2011_review_QC}. At the quantum critical point (QCP), continuous phase transitions are governed by quantum fluctuations, leading to the emergence of exotic material properties, including non-Fermi-liquid (NFL) behavior~\cite{2001_review_NFL,2001_Howdo,2007_review_FL_instabilities}, heavy fermion behavior~\cite{1998_nature_HF,2000_nature_HF,2003_nature_HF,2008_np_HF}, and unconventional superconductivity~\cite{2010_review_puzzle,2010_science_review_HF,2012_np_review_Fe,2015_Fe_review,2015_review_copper,2024_nature_CsCrSb}. The breakdown of Landau's Fermi-liquid theory near QCPs underscores the significance of electron correlations and scale-invariant quantum fluctuations, challenging traditional notions of ordering and quasiparticle coherence~\cite{2004_prb_QC,2012_UQCP}.

Quantum criticality (QC) in materials without fine-tuning of parameters can predominantly emerge from geometric frustrations~\cite{2010-nature-review,2017rmpqsl,2017phyreportFrustrated,2018_frustraion_QC}. Consequently, the suppression of conventional order by quantum fluctuations brings about unusual ground states, such as quantum spin liquids or valence-bond solids~\cite{2006_prl_Pr2Ir2O7,2007prlNaIrO,2008_np_organic,2021_nature_Fractionalized,2014_nm_QCP,2016_nature_YbMgGaO4,
2019_np_Ce2Zr2O7,2021_prx_NaYbSe2,2010_np_VBS,2012_nm_LiZn2Mo3O8}. In contrast, interaction-driven frustration in square lattices, where geometric frustration is inherently absent, seldom results in genuine QCPs, as these materials typically exhibit long-range magnetic order (LRMO) at lower temperatures~\cite{2004_Pb2VO(PO4)2,2008_BaCdVO(PO4)2,2009_PVP,2010_PbZn(PO4)2,2011_SrZnVO(PO4)2,2005_CuClLaNd2O7,2006_CBLNO,2011_NaVOPO4F,2022_prm_Sr2NiO3ClF}.
At present, the interplay between interaction-driven frustration and QC in metallic square lattices remains insufficiently explored.

In this Letter, we report a new intermetallic compound, ThCr$_2$Ge$_2$C, hosting a Cr$_2$C square-planar lattice, where competing nearest-neighbor (NN) $J_1$ and next-nearest-neighbor (NNN) $J_2$ interactions with $J_2/J_1 \sim -0.5$ stabilize QCPs at zero-field and ambient pressure. Our experimental investigations reveal that ThCr$_2$Ge$_2$C exhibits NFL behavior in its resistivity and specific heat, along with the absence of LRMO down to~70~mK. Under high pressures, which alleviate the underlying magnetic frustration, the NFL behavior disappears experimentally, along with the stabilization of magnetically ordered phases as revealed by the calculations. These findings indicate that ThCr$_2$Ge$_2$C could serve as a model system for studying QC and possible spin-nematic phase in metallic frustrated magnets.

\begin{figure}[t]
	\includegraphics[width=8.8cm]{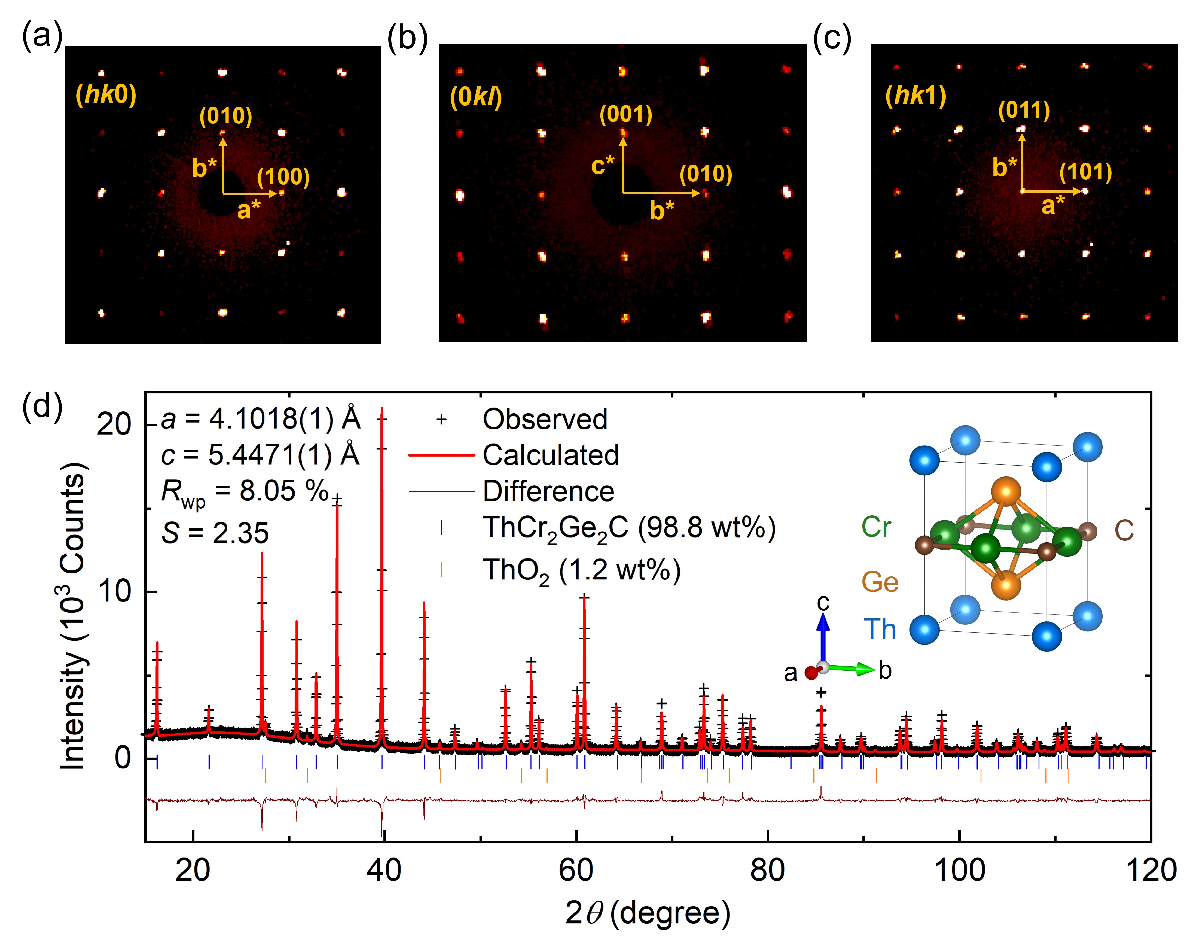}
    \centering
	\caption{(a-c) Reconstructed  ThCr$_2$Ge$_2$C single-crystal XRD patterns of ($hk$0), (0$kl$) and ($hk$1) reflection planes, respectively. (d)~Powder XRD at room temperature and its Rietveld refinement profile of ThCr$_2$Ge$_2$C. The inset shows the crystal structure of ThCr$_2$Ge$_2$C.}
	\label{fig1_XRD}
\end{figure}

Single and polycrystalline crystals of ThCr$_2$Ge$_2$C were synthesized using the arc-melting technique. Additional details regarding the experimental setup and theoretical calculations are provided in the Supplemental Materials (SM)~\cite{SM}. The single crystals were characterized via single-crystal X-ray diffractions (XRD) and energy dispersive X-ray spectroscopy (Fig.~\ref{fig1_XRD}(a-c) and Fig.~S1). The polycrystalline sample was analyzed by XRD at room temperature~(Fig.~\ref{fig1_XRD}(d)). The XRD results show that ThCr$_2$Ge$_2$C crystallizes in a tetragonal lattice with space group $P$4/$mmm$ (Tables S1 and S2), where the Cr$_2$C square lattice adopts an anticonfiguration to the CuO$_2$ square net in cuprate superconductors (inset of Fig. 1(d))~\cite{2015_review_copper}.

\begin{figure}[htp]
	\includegraphics[width=8.8cm]{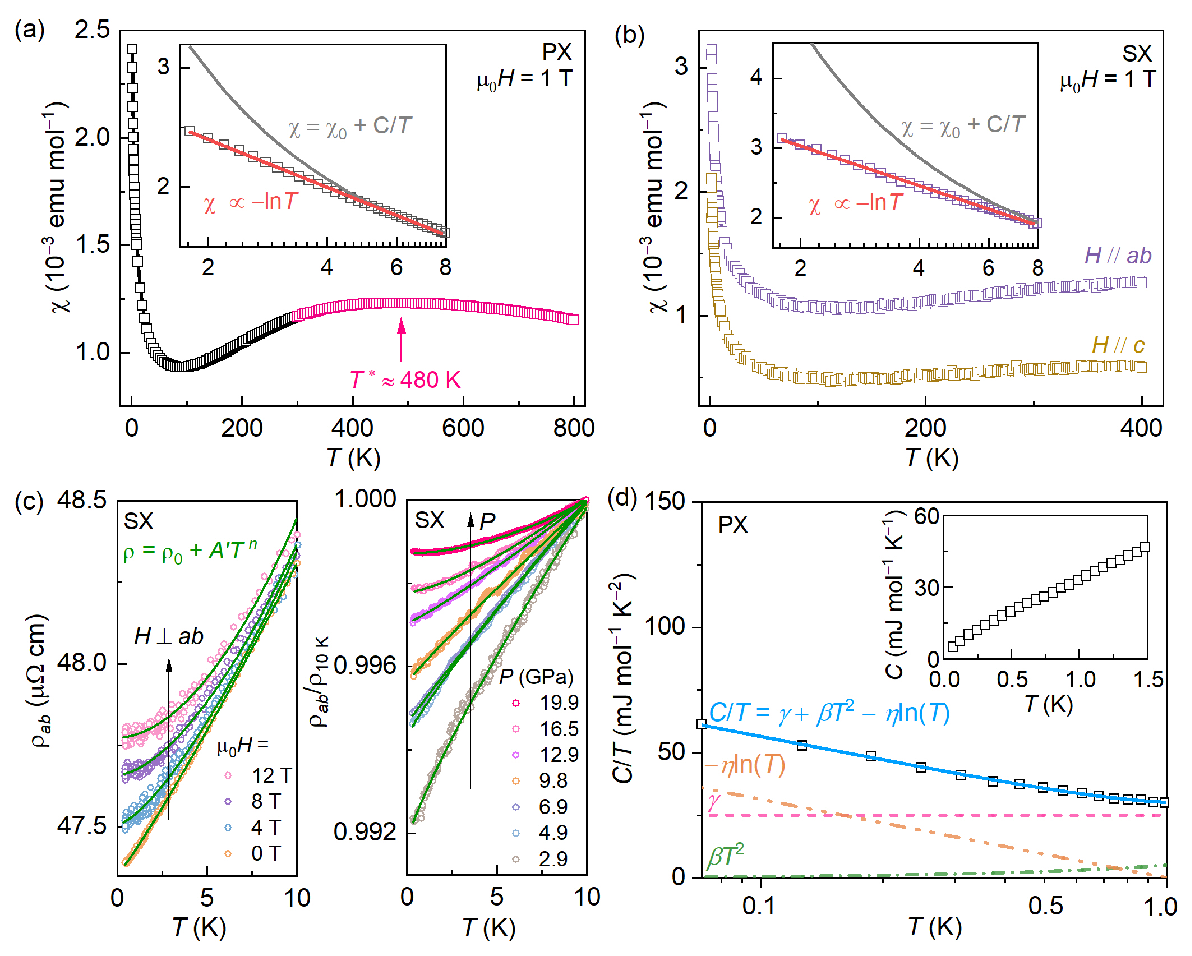}
    \centering
\caption{(a) Temperature dependence of magnetic susceptibility, $\chi(T)$, for the polycrystalline ThCr$_2$Ge$_2$C sample in a 1~T magnetic field. The data points in black were collected from 1.8~K to 300~K, and those in pink from 300~K to 800~K. The inset shows low-temperature $\chi(T)$. (b) $\chi(T)$ for a single crystal in a 1~T magnetic field, with the inset showing low-temperature $\chi(T)$. (c) Left panel is temperature dependence of the $ab$-plane electrical resistivity, $\rho_{ab}(T)$, for the single crystal under various magnetic fields applied perpendicular to the $ab$-plane. Right panel shows $\rho_{ab}/\rho_{\rm 10 K}$ under different pressures. The green lines show the low-temperature resistivity fitted with $\rho = \rho_0 + A'T^n$. (d) Plot of $C/T$ versus $T$, with the blue line fitting $C/T = \gamma + \beta T^2 - \eta \ln(T)$. The dashed lines represent the contribution from each term. The inset shows the specific heat, $C(T)$, for the polycrystalline sample at low temperatures. PX denotes polycrystalline samples and SX denotes single crystals.}
	\label{fig2_PP}
\end{figure}

Fig.~\ref{fig2_PP}(a) shows the temperature-dependence magnetic susceptibility $\chi(T)$ of the polycrystalline sample. As observed, a broad hump appears with its maximum at $T^{\ast} \approx 480$~K. This feature, distinct from a sharp cusp or kink for antiferromagnetic (AFM) transition, is reminiscent of spin frustration, low-dimensional magnetism, or short-range correlations~\cite{2018_npj,2009EPLmag}. The high-temperature data cannot be described by the extended Curie-Weiss formula, suggesting that AFM correlations persist up to 800 K. As a result, the Curie-Weiss temperature ($\theta_{\mathrm{CW}}$) is at least above 480~K, and the magnetic frustration index~\cite{1994_findex}, $f = |\theta_{\mathrm{CW}}|/ T_{\mathrm{N}}$ $>$~6850 ($T_{\mathrm{N}}$ cannot be detected down to 70 mK, see below), turns out to be extraordinarily large, indicating extremely strong magnetic frustration~\cite{2010-nature-review}. 

The $\chi(T)$ of the single crystal shows a similar increasing trend to that of the polycrystalline sample above 100~K, demonstrating that these magnetic properties are intrinsic (Fig.~\ref{fig2_PP}(b)). The low-temperature $\chi(T)$ under a 10~Oe magnetic field shows no bifurcation in the zero-field-cooled (ZFC) and field-cooled (FC) data down to 2~K. Additionally, the ac susceptibility data show neither cusp-like anomaly nor frequency dependence, ruling out the possibility of a spin-glass state in the system~(Fig.~S2). At low temperatures,  both single and polycrystalline samples exhibit a $-\ln T$ dependence below 8~K, deviating from the Curie law~(insets of Figs.~\ref{fig2_PP}(a,b)). This behavior is indicative of QC, due to strong quantum spin fluctuations~\cite{2001_review_NFL,2002_prb_Kondo,2006_prl_Pr2Ir2O7}. 

The temperature dependence of electrical resistivity for the single crystal was measured under various magnetic fields and pressures (Fig.~\ref{fig2_PP}(c) and Figs.~S3(a,b)). The low-temperature data (45~mK $< T <$ 10~K) at zero-field and ambient pressure were fitted using the formula $\rho~=~\rho_0~+~A'T^n$, yielding $\rho_0 = 47$~$\mu\Omega$~cm and $n = 1.11$. The value of $n$ is significantly smaller than 2.0, suggesting breakdown of the Fermi--liquid scenario. The NFL behavior may be induced by quantum critical fluctuations~\cite{2001_review_NFL, 1998_nature_HF, 2003_nature_HF}. Similar fits to the data under magnetic fields and high pressures show an increase in $n$~(Fig.~S3(c)). Magnetic field or pressure can tune the quantum critical behavior, suppressing quantum fluctuations in the NFL state and driving the system to a Fermi liquid~\cite{1996_CeCu6-xAsx,2002_prl_YbRh2Si2,2003_prl_CeCoIn5, 2008_Field_FL,2024_prl_CeRh2As2}. Assuming a Fermi--liquid scenario in the low-temperature limit, the data can alternatively be fitted using the formula $\rho~=~\rho_0~+~AT^2$. The coefficient $A$ under magnetic fields or pressures exhibits divergence as $\mu_0 H \rightarrow 0$ T or $P \rightarrow 0$ GPa (Fig.~S3(d-f)), corroborating the presence of QC at zero field and ambient pressure~\cite{2008_Field_FL,2024_prl_CeRh2As2,2024_nature_CsCrSb}.

The temperature dependence of the specific heat for the polycrystalline sample shows no obvious anomaly, indicating the absence of LRMO or short-range spin freezing down to 70~mK~(inset of Fig.~\ref{fig2_PP}(d) and Fig.~S4). Data in the temperature range of 70~mK to 1~K were well described by the formula $C/T = \gamma + \beta T^2 - \eta \ln(T)$~(Fig.~\ref{fig2_PP}(d)). The data fitting yields $\gamma = 24.98$~mJ~K$^{-2}$~mol$^{-1}$, which could be due to quasiparticle excitations in the residual ``cold region" of the Fermi surface and/or other excitations like spinons~\cite{2008_np_organic,2016_nature_YbMgGaO4,1999_prl_cold_region}. Notably, $C/T$ exhibits a $-\ln T$ dependence, which is not due to Schottky anomaly (inset of Fig.~S4), but consistent with the quantum critical behavior observed in $\chi(T)$ and $\rho(T)$~\cite{1995_QCP, 2001_review_NFL, 1996_CeCu6-xAsx}.

\begin{figure}[t]
	\includegraphics[width=8.8cm]{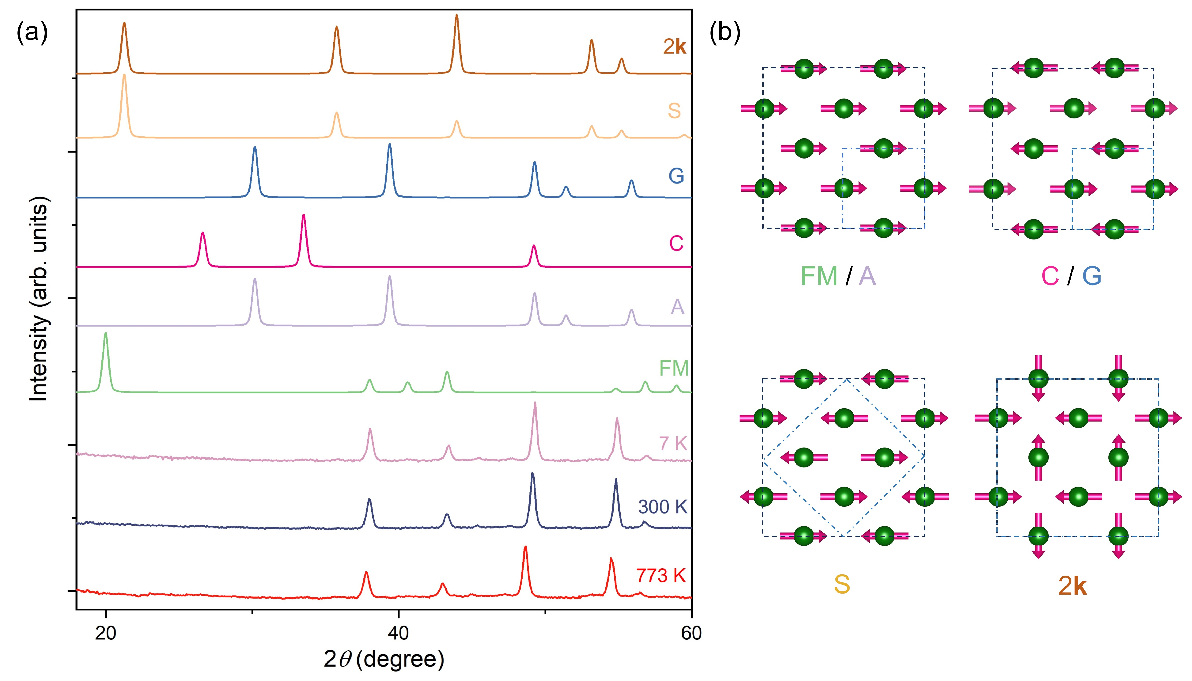}
    \centering
	\caption{(a) Simulated neutron powder diffraction (NPD) patterns for different magnetic structures, compared with NPD data for ThCr$_2$Ge$_2$C at 7~K, 300~K, and 773~K. (b) Diagrams of magnetic structures with Cr spins aligned within the $ab$-plane: FM, C (ferromagnetic interlayer coupling), A, G, S, 2\textbf{k} (antiferromagnetic interlayer coupling). The blue dashed-dotted squares denote the magnetic unit cell for different magnetic structures.}
	\label{fig3_npd}
\end{figure}

Neutron powder diffraction (NPD) experiments were conducted to confirm the absence of LRMO (Fig.~S5 and Table~S2). Fig.~\ref{fig3_npd} compares the simulated NPD patterns of the most likely magnetic structures (with Cr spins aligned within the $ab$-plane) to the experimental data at 7~K, 300~K, and 773~K~\cite{2018_UCr2Si2C, 2023_ThCrSiC, 2006_prb_hp_BaFeSeO, 2022_prm_BaFeSO}. No magnetic Bragg peaks were observed, only nuclear peaks, indicating the absence of LRMO down to 7~K. A comparison with Cr spins aligned along the $c$-axis yields the same result (Fig.~S6). No additional peaks were found when comparing the 7~K and 300~K data with the 773~K data (well beyond $T^{\ast} \approx 480$~K), further supporting the absence of LRMO.

\begin{figure*}[t]
	\includegraphics[width=18cm]{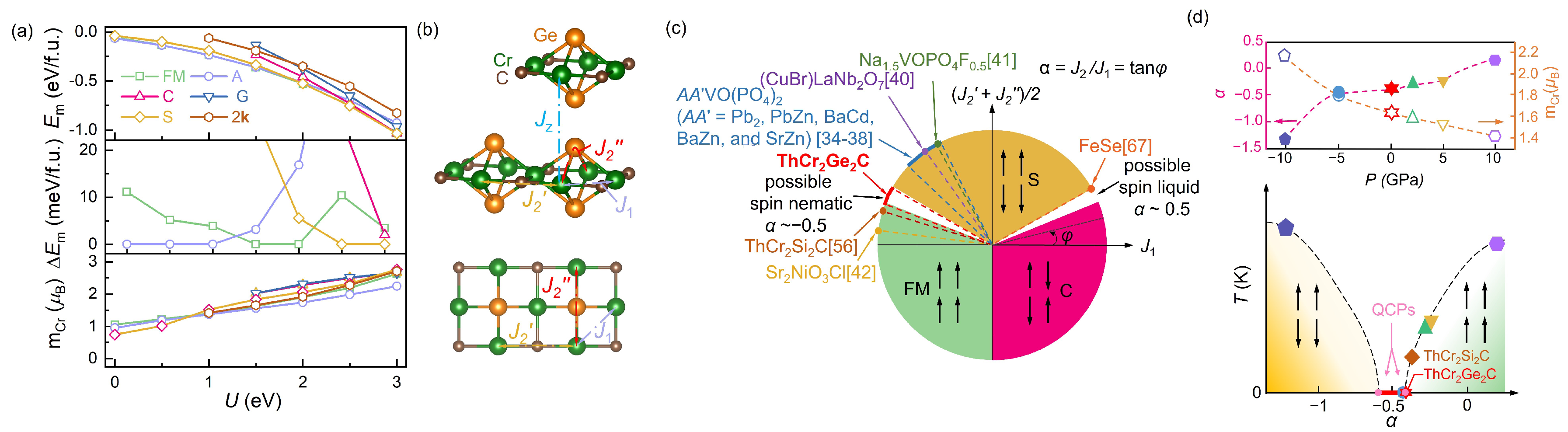}
    \centering
	\caption{(a) Calculation of the energy of ThCr$_2$Ge$_2$C for different magnetic configurations and their corresponding magnetic moment of Cr. $E_{\rm m}$ represents the energy relative to non-magnetic configuration. $\Delta E_{\rm m}$ is the energy difference between the magnetic ground state and other magnetic structures at each $U$ value. (b) The magnetic exchange interactions between the Cr spins. (c) Modified phase diagram in the two-dimensional $J_1 - J_2$ model following Refs.~\cite{2009_PVP,2017phyreportFrustrated}. (d) The upper panel shows the pressure dependence of $\alpha$ (left axis) and magnetic moment (right axis) from DFT calculations, with the dashed lines as a guide to the eye. The lower panel presents the putative $T-\alpha$ phase diagram.}
	\label{fig4_dft}
\end{figure*}

In square-lattice systems, where geometric frustration is absent, LRMO is typically suppressed by interaction frustration from competing NN and NNN exchange couplings~\cite{2004_epjb-Hmodel,2006_prl_Nematic,2010PRBJ1J2,2019-npj-review}. To reveal the magnetic frustration, we performed density functional theory (DFT)-based first-principles calculations~(Fig.~S7 and Table~S3). Fig.~\ref{fig4_dft}(a) shows the calculated energy relative to the non-magnetic (NM) state $E_\text{m}$ as a function of $U$ for different magnetic configurations. The magnetic ground state evolves from A to FM, and then to S as the $U$ value increases, with small energy differences $\Delta E_{\rm m}$ between the  ground state and other spin orders at each $U$ value, indicating instability of the LRMO. To estimate interaction parameters in ThCr$_2$Ge$_2$C, we performed the constrained random-phase approximation (cRPA) calculations. The results give $U = 1.25$~eV and $J/U = 0.36$, close to those in the kagome metal CsCr$_3$Sb$_5$\cite{2025_nc_CsCr3Sb5}. Thus, a $U$ value of about $\sim$1~eV should be included to account for electron-electron interactions. 

\begin{table}[t]
\centering
\fontsize{9}{15}\selectfont
\caption{The spin exchange interaction constants, $J_1$, $J'_2$, $J''_2$ and $J_z$ in ThCr$_2$Ge$_2$C. For $U = 1$~eV, $J'_2$ and $J''_2$ cannot be given separately. Therefore, we only provide an average value. The last column $\alpha$ is defined by $\alpha = (J'_2 + J''_2) / 2J_1 $. }
   \begin{tabular}{p{1.5cm}<{\centering}|p{1.5cm}<{\centering}p{1.5cm}<{\centering}p{1.5cm}<{\centering}p{1.4cm}<{\centering}|p{1.4cm}<{\centering}}
\hline\hline
\multirow{2}{*}{$U$ (eV)} & \multicolumn{4}{c|}{$J_i$ (meV/$S^2$)}                                                         & \multirow{2}{*}{$\alpha$} \\ \cline{2-5}
                          & \multicolumn{1}{c|}{$J_1$}  & \multicolumn{1}{c|}{$J'_2$} & \multicolumn{1}{c|}{$J''_2$} & $J_z$ &                           \\ \hline
1                         & \multicolumn{1}{c|}{$-$74.94} & \multicolumn{2}{c|}{31.41}                               & 0.72  & $-$0.42                     \\ \hline
1.5                       & \multicolumn{1}{c|}{$-$27.99} & \multicolumn{1}{c|}{36.37} & \multicolumn{1}{c|}{$-$14.38} & $-$0.79 & $-$0.39                     \\ \hline
2                          & \multicolumn{1}{c|}{$-$17.67} & \multicolumn{1}{c|}{43.73} & \multicolumn{1}{c|}{$-$23.24} & $-$4.22 & $-$0.58  \\ \hline\hline                
\end{tabular}
    \label{magnetic structure}
\end{table}

The exchange interactions between the Cr atoms can be calculated using a $J_1$-$J_2'$-$J_2''$-$J_z$ Heisenberg model (including a $J_3$ term yields nearly identical results, as detailed in the SM~\cite{SM}). The effective magnetic interactions between Cr spins can be modeled by: (i) the NN coupling $J_1$ with direct Cr-Cr exchange, $\sim$70$^\circ$ Cr-Ge-Cr superexchange, and 90$^\circ$ Cr-C-Cr superexchange, (ii) the NNN $J_2'$ with 180$^\circ$ Cr-C-Cr superexchange, (iii) the NNN $J''_2$ with $\sim$108$^\circ$ Cr-Ge-Cr superexchange, and (iv) the interlayer NN coupling $J_z$, as shown in Fig.~\ref{fig4_dft}(b). Note that RKKY interactions, if not negligible, could also be included. The energies of the FM, A, C, G, S, and 2\textbf{k} magnetic states can be expressed as follows:
\begin{equation*}
    \left\{
    \begin{array}{l}
        E_{\rm FM} - E_0 = 4J_1 + 2J'_2 + 2J''_2 + 2J_z, \\
        E_{\rm A} - E_0 = 4J_1 + 2J'_2 + 2J''_2 - 2J_z, \\
        E_{\rm C} - E_0 = -4J_1 + 2J'_2 + 2J''_2 + 2J_z, \\
        E_{\rm G} - E_0 = -4J_1 + 2J'_2 + 2J''_2 - 2J_z, \\
        E_{\rm S} - E_0 = -2J'_2 - 2J''_2 - 2J_z, \\
        E_{\rm 2\textbf{k}} - E_0 = 2J'_2 - 2J''_2 - 2J_z, \\
    \end{array}
    \right.
\end{equation*}
where $E_{\rm FM}$, $E_{\rm A}$, $E_{\rm C}$, $E_{\rm G}$, $E_{\rm S}$, and $E_{2\textbf{k}}$ denote the magnetic energy of those magnetic structures, respectively, and $E_0$ is an energy offset. The spin-exchange parameters, $J_1$, $J'_2$, $J''_2$ and $J_z$ can be derived for $U \ge 1$~eV (for $U = 0$ and $0.5$~eV, these values cannot be obtained, as the 2\textbf{k} state converges to a NM state), as listed in Table~\ref{magnetic structure}. $J_1$ values are predominantly FM, with $J'_2$ always being AFM, and $J''_2$ being FM, all of which are consistent with the Goodenough-Kanamori rules~\cite{1955_LaMMnO3,1958_LaSrCoO3,1957_FeCoO}. $J_z$ is about one order of magnitude smaller than $J_1$, $J'_2$, and $J''_2$, suggesting a quasi-two-dimensional~(2D) magnetism in ThCr$_2$Ge$_2$C. 


In a 2D scenario, there exist several phases as the ratio $\alpha =J_2/J_1 =\tan\varphi$ varies. Quantum fluctuations disrupt LRMO and result in the formation of two quantum critical regions at the boundaries of the ordered phases (Fig.~\ref{fig4_dft}(c))~\cite{2009_PVP,2010PRBJ1J2,2017phyreportFrustrated}. A relevant example is the iron-chalcogenide superconductor FeSe, where quantum fluctuations associated with strongly frustrated interactions are identified as a potential trigger for the nematic quantum paramagnetic phase, interpolating between C- and S-type magnetic instabilities at $\alpha \sim 0.6$~\cite{2015_np_FeSe,2016_nc_FeSe,2016_nm_FeSe}. For ThCr$_2$Ge$_2$C, the Cr$_2$C planes form a Lieb-like square-planar lattice, consisting of two Cr sites. One site has $J'_2$ along the $a$ axis and $J''_2$ along the $b$ axis, while the other is the opposite. Therefore, the effective $J_2$ should be ($J'_2 + J''_2$)/2, after taking the coordination number of Cr into account.

The result of $J_1 < 0$ and $(J'_2 + J''_2)/2 > 0$~(Table~\ref{magnetic structure}) indicates a spin-frustrated state in ThCr$_2$Ge$_2$C, which is commonly seen in square-lattice quantum magnets~\cite{2004_Pb2VO(PO4)2,2008_BaCdVO(PO4)2,2009_PVP,2010_PbZn(PO4)2,2011_SrZnVO(PO4)2,2005_CuClLaNd2O7,2006_CBLNO,2011_NaVOPO4F,2022_prm_Sr2NiO3ClF}. Remarkably, the value of $\alpha = (J'_2 + J''_2) /2J_1$  is close to the maximal frustration, which could give rise to the possible spin nematic state as well as the substantial quantum fluctuations. Combined with experimental observations of NFL transport property, as well as logarithmic divergences in thermodynamic and magnetic susceptibilities, we propose that ThCr$_2$Ge$_2$C could be a gapless quantum magnet near a QCP, induced by strong magnetic frustration~(Fig.~\ref{fig4_dft}(d)). 

To further address the above proposal, we performed DFT calculations to investigate the pressure effects on ThCr$_2$Ge$_2$C at $U = 1.5$~eV (Fig.~S9 and Table~S5). The result indicates that the magnetic ground state evolves from an S-type structure under negative pressure to a FM/A-type structure under positive pressure. Simultaneously, the value of $\alpha$ increases with pressure, showing a small change in the low-pressure range ($\pm$5~GPa) and more significant changes at higher pressures ($\pm$10~GPa)~(Fig.~\ref{fig4_dft}(d)). Notably, the magnetic moment does not disappear in the quantum-critical regime, distinct from ordinary magnetic QC. For the isoelectronic analogue ThCr$_2$Si$_2$C, both NPD data and DFT calculations indicate an A-type magnetic ground state with in-plane FM ordering~\cite{2023_ThCrSiC}. The substitution of Si for Ge induces a positive chemical pressure effect, consistent with the positive pressure effect calculated for ThCr$_2$Ge$_2$C. These findings strongly suggest that pressure may shift the system away from the quantum critical region and into a magnetically ordered state.

In summary, we report the discovery of ThCr$_2$Ge$_2$C, a metallic quantum magnet with a square-planar Cr lattice. Neither LRMO nor short-range spin freezing is detected by experiments down to 70~mK. DFT calculations reveal that the competition between $J_1$ and $J_2$ drives the QC. Thus, ThCr$_2$Ge$_2$C could serve as a prototype for studying the role of competing interactions in QC. In-depth investigations utilizing techniques, such as muon spin relaxation, neutron scattering, and nuclear magnetic resonance, are expected to further address the QC  and possible spin nematic phase in the future.

We acknowledge Jun Zhao for helpful discussions. This work was supported by the National Key Research and Development Program of China (2023YFA1406101, 2022YFA1403202), and the Natural Science Foundation of Shandong Province, China (ZR2023MA028).
\clearpage
\begin{widetext}
\begin{center}
    \normalsize\textbf{Supplemental Materials: Quantum Criticality by Interaction Frustration in a Metallic Square-Planar Lattice} \\
\end{center}
\renewcommand{\thefigure}{S\arabic{figure}}
\renewcommand{\thetable}{S\arabic{table}}
\setcounter{figure}{0}
\setcounter{table}{0}

\section{Contents}
\begin{spacing}{1.8}

Experimental and computational methods

Fig.~\hyperref[figS1_edx]{S1}: Characterizations of ThCr$_2$Ge$_2$C single crystals.

Fig.~\hyperref[figS2_AC]{S2}: Low-temperature dc and ac magnetic susceptibility of ThCr$_2$Ge$_2$C single crystals.

Fig.~\hyperref[figS3_R]{S3}: Temperature dependence of electrical resistivity for the single crystal of ThCr$_2$Ge$_2$C.

Fig.~\hyperref[figS4_C]{S4}: Specific heat for the polycrystalline sample of ThCr$_2$Ge$_2$C.

Fig.~\hyperref[figS5_NPD]{S5}: Neutron powder diffraction patterns of ThCr$_2$Ge$_2$C at 7~K, 300~K and 773~K.

Fig.~\hyperref[figS6_compare]{S6}: Comparison of simulated neutron diffraction patterns with experimental data.

Fig.~\hyperref[figS7_dosband]{S7}: Orbital-projected band structures and corresponding density of states of ThCr$_2$Ge$_2$C.

Fig.~\hyperref[figS8_J3]{S8}: Diagrams of the diagonal double stripe magnetic structure and the magnetic exchange interactions.  

Fig.~\hyperref[figS9_dftHP]{S9}: DFT calculation results for ThCr$_2$Ge$_2$C under positive and negative pressures.  

Table~\hyperref[Crystal]{S1}: Crystallographic data for the single crystal ThCr$_2$Ge$_2$C at 298~K.  

Table~\hyperref[NPD]{S2}: Refined results for ThCr$_2$Ge$_2$C from X-ray and neutron powder diffractions at various temperatures.  

Table~\hyperref[dft]{S3}: DFT calculation results for ThCr$_2$Ge$_2$C.  

Table~\hyperref[J3]{S4}: Calculated spin exchange interactions and the $\alpha$ in ThCr$_2$Ge$_2$C with $J_3$ term is included.  

Table~\hyperref[dftHP]{S5}: DFT calculation results for ThCr$_2$Ge$_2$C under positive and negative pressures.  
\end{spacing}
\newpage

\section{Experimental and computational methods}
\label{sec:experimental}
\begin{spacing}{1.11}
The polycrystalline samples of ThCr$_2$Ge$_2$C were synthesized using the powder of Th, Cr (99.95\%), Ge (99.999\%) and graphite (99.95\%) as starting materials. The preparation of thorium metal was mentioned at Ref.~\cite{2016_Th}. The stoichiometric mixture of the materials was sufficiently mixed in an agate mortar and cold-pressed into a pellet. All these procedures were carried out in a glovebox filled with high-purity argon to prevent oxidation. The pellet was then melted in a water-cooled copper hearth by using an arc melting furnace. The sample was melted serval times and flipped after each melting to ensure homogeneity. The arc-melted ingot was thoroughly ground, placed in an alumina crucible, sealed in an evacuated silica tube, and then heated to 1000 $^{\circ}$C for two weeks. The single crystals of ThCr$_2$Ge$_2$C were carefully extracted from the ingot obtained through the arc melting process.

Single-crystal X-ray diffraction (XRD) was carried on a Bruker D8 Venture diffractometer with Mo-$K_{\alpha}$ radiation. The data reduction was done using the commercial software package APEX4. The reconstructed images in the reciprocal space from the raw frames were produced using the reciprocal unit vectors of the tetragonal lattice by the software CrysAlis$^{\rm Pro}$ (CrysAlis Pro v.171.40.53, Rigaku Oxford Diffraction). The XRD for polycrystalline sample was performed at room temperature using a PANalytical X-ray diffractometer with Cu-$K_{\alpha1}$ radiation. The chemical compositions of single crystals were confirmed by energy dispersive X-ray (EDX) spectroscopy (Oxford Instruments X-Max) equipped in a scanning electron microscope (SEM, Hitachi S-3700N). Neutron powder diffraction (NPD) measurements were carried out on the high-resolution neutron diffractometer at the Key Laboratory of Neutron Physics, Institute of Nuclear Physics and Chemistry, China Academy of Engineering Physics. The wavelength of the neutron was $\lambda$ = 1.8846~\r{A}. The crystal structure was refined by using the FullProf suite, with magnetic structures were analyzed with SARAH software~\cite{1993rodriguez.fullprof,2000_sarah}. 

A standard four-terminal method was employed to measure the temperature-dependent resistivity.  Resistivity measurements above 2.0~GPa were performed as described in Ref.~\cite{2024_np_La327}, with details on the setup and methodology. The heat capacity was measured utilizing the standard relaxation method, using a Quantum Design physical property measurement system (PPMS) with dilution refrigerator inserts down to 70 mK. The magnetic properties were measured by a Quantum Design magnetic property measurement system (MPMS-3). The measurement was allowed at temperatures up to 800 K by employing a high-temperature option. 

The first-principles calculations were done within the generalized gradient approximation (GGA) by using the Vienna Ab-initio Simulation Package (VASP)~\cite{gga,vasp}. We used the experimental crystal structure parameters for the initial relaxation. The forces were minimized to less than 0.0001 eV/\r{A} in the relaxation. A plane-wave basis energy cutoff of 500 eV was employed alongside a 20$\times$20$\times$15 $\Gamma$-centered K-mesh for DOS calculations. The Coulomb and exchange parameters, $U$ and $J$, were introduced by using the GGA + $U$ calculations, where the parameters $U$ and $J$ are not independent and the difference ($U_{\mathrm{eff}}$ = $U - J$) is meaningful. Considering NPD experimental results for ThCr$_2$Ge$_2$C, along with studies on other Th-based compounds such as ThB$_2$C~\cite{1989_ThB2C}, ThCr$_2$Si$_2$~\cite{2012_ThCr2Si2}, ThCr$_2$Si$_2$C~\cite{2023_ThCrSiC}, all showing that Th does not carry a magnetic moment. Given that standard DFT calculations often overestimate the hybridization between $f$-electrons and conduction electrons, we fixed the $U_{\mathrm{eff}}$  at 11.0~eV for the Th 5$f$ orbitals to mitigate this effect~\cite{2023_ThCrSiC,2023_ThCrAsN}.

To assess the relevance of long-range interactions, we evaluated the magnitude of the long-range interaction $J_3$ via calculating an additional magnetic structure, the diagonal double stripe (DDS) (Fig.~S8), at $U$ = 1.5, 2~eV. The energy ($E_m$) of the DDS magnetic state, relative to the non-magnetic configuration, is $–$278.81~meV/f.u. and $–$489.31~meV/f.u., respectively. Then the energies of the FM, A, C, G, S, 2\textbf{k} and DDS magnetic states can be expressed as follows:
\begin{equation*}
    \left\{
    \begin{array}{l}
        E_{\rm FM} - E_0 = 4J_1 + 2J'_2 + 2J''_2 + 4J_3 + 2J_z, \\
        E_{\rm A} - E_0 = 4J_1 + 2J'_2 + 2J''_2 + 4J_3 - 2J_z, \\
        E_{\rm C} - E_0 = -4J_1 + 2J'_2 + 2J''_2 + 4J_3 + 2J_z, \\
        E_{\rm G} - E_0 = -4J_1 + 2J'_2 + 2J''_2 + 4J_3 - 2J_z, \\
        E_{\rm S} - E_0 = -2J'_2 - 2J''_2 + 4J_3 - 2J_z, \\
        E_{\rm 2\textbf{k}} - E_0 = 2J'_2 - 2J''_2 - 4J_3 - 2J_z, \\
        E_{\rm DDS} - E_0 = - 4J_3 - 2J_z, \\
    \end{array}
    \right.
\end{equation*}
The spin-exchange parameters derived are summarized in Table~S4. The results show that $J_3$ is about one order of magnitude smaller than $J_1$, $J'_2$, and $J''_2$, and that the $\alpha$ values remain nearly identical when $J_3$ is excluded.

\end{spacing}

\begin{figure*}[htbp]
	\includegraphics[width=16cm]{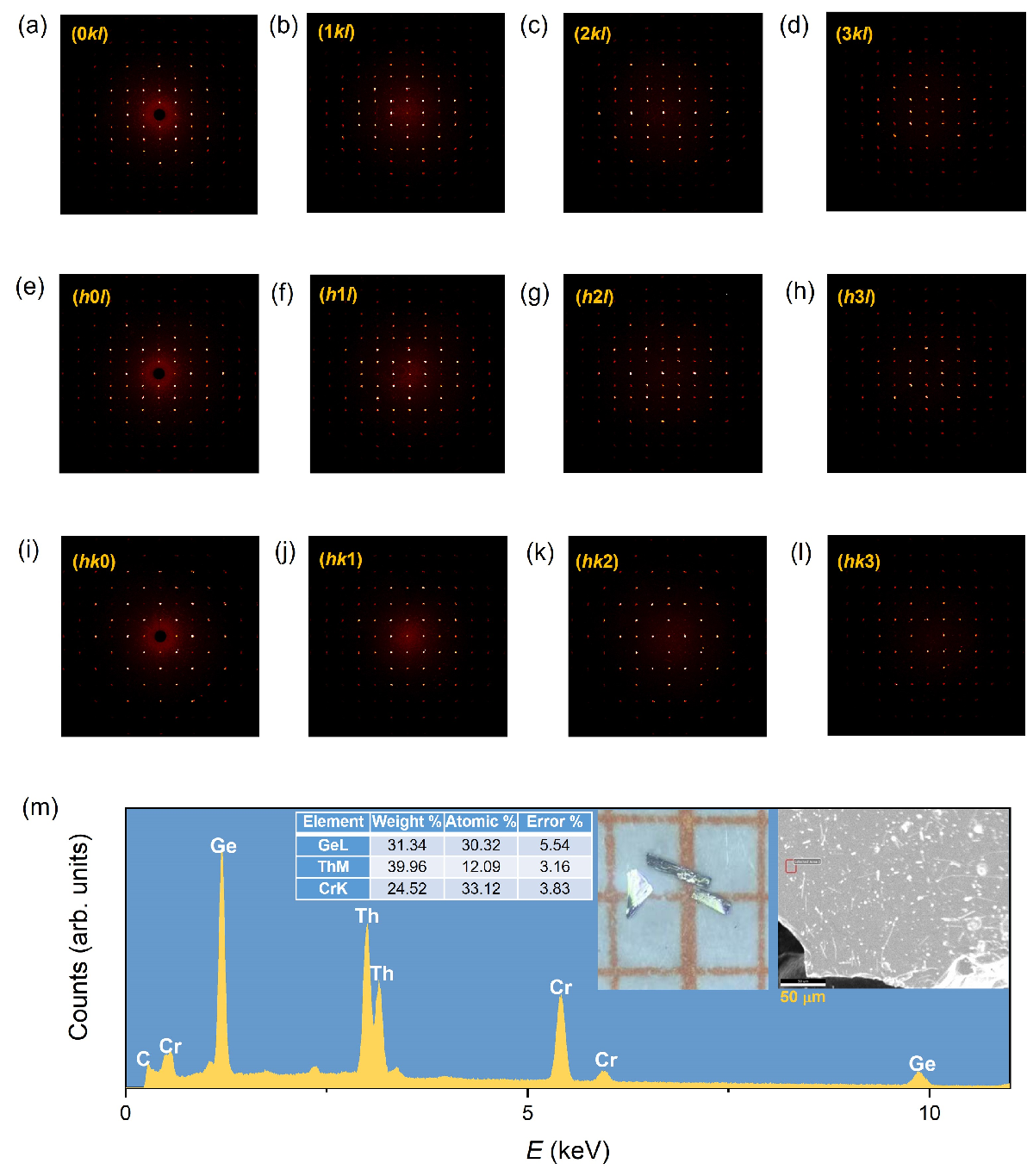}
    \centering
	\caption{(a-l) Reconstructed specific ($hkl$) planes of the ThCr$_2$Ge$_2$C single crystal at 298~K. (m) The typical EDS spectrum of the ThCr$_2$Ge$_2$C single crystal. The top-left inset shows the optical photograph, and the top-right inset shows an SEM image. The atomic ratios of Th, Cr, and Ge are in good agreement with the stoichiometry.}
	\label{figS1_edx}
\end{figure*}

\begin{figure*}[htbp]
	\includegraphics[width=10cm]{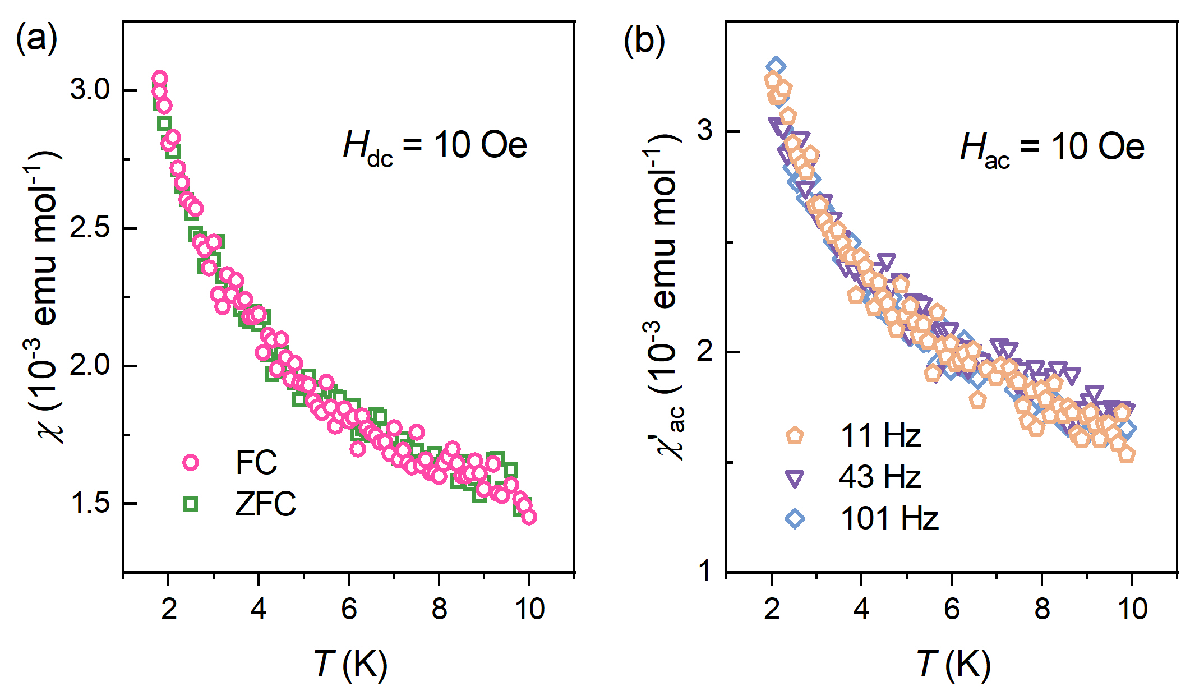}
    \centering
	\caption{(a) Low-temperature magnetic susceptibility of the ThCr$_2$Ge$_2$C single crystal with zero-field-cooling (ZFC) and field-cooling (FC) protocols under a 10 Oe magnetic field. (b) Real part of ac magnetic susceptibility at different oscillating frequencies under zero dc field.}
	\label{figS2_AC}
\end{figure*}

\begin{figure*}[htbp]
	\includegraphics[width=14cm]{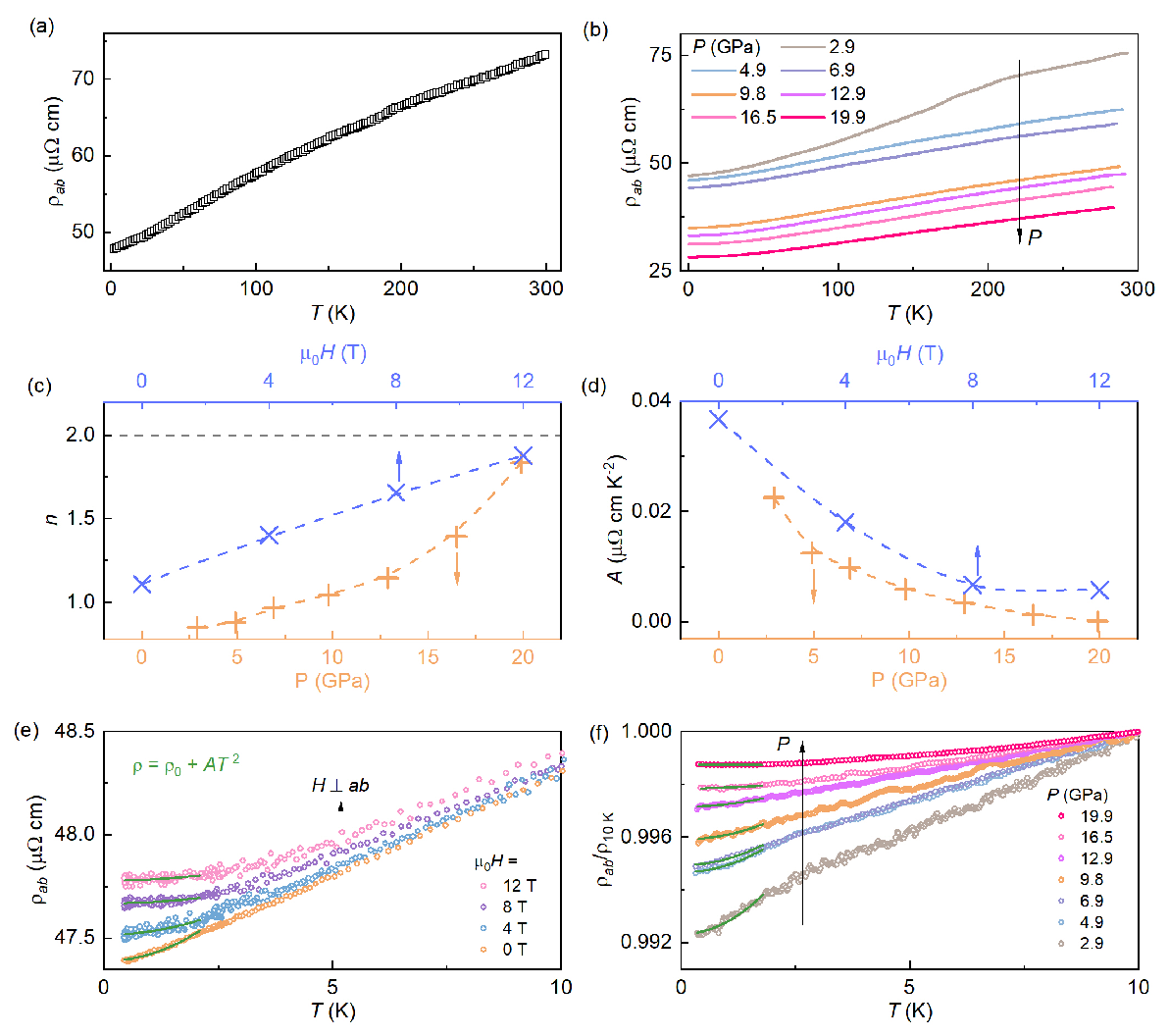}
    \centering
	\caption{(a) Temperature dependence of $ab$ electrical resistivity, $\rho(T)$, for the ThCr$_2$Ge$_2$C single crystal. (b) $\rho_{ab}(T)$ of the single crystal under various pressures. (c) Power $n$ as a function of magnetic field (upper axis) or pressure (lower axis). (d) Coefficients $A$ as functions of magnetic field (upper axis) or pressure (lower axis). The dashed lines serve as a guide to the eye. (e, f) Fits of the $ab$-plane electrical resistivity under various magnetic fields applied perpendicular to the $ab$-plane and under different pressures, respectively, using the formula $\rho = \rho_0 + AT^2$.}
	\label{figS3_R}
\end{figure*}

\begin{figure*}[htbp]
	\includegraphics[width=8cm]{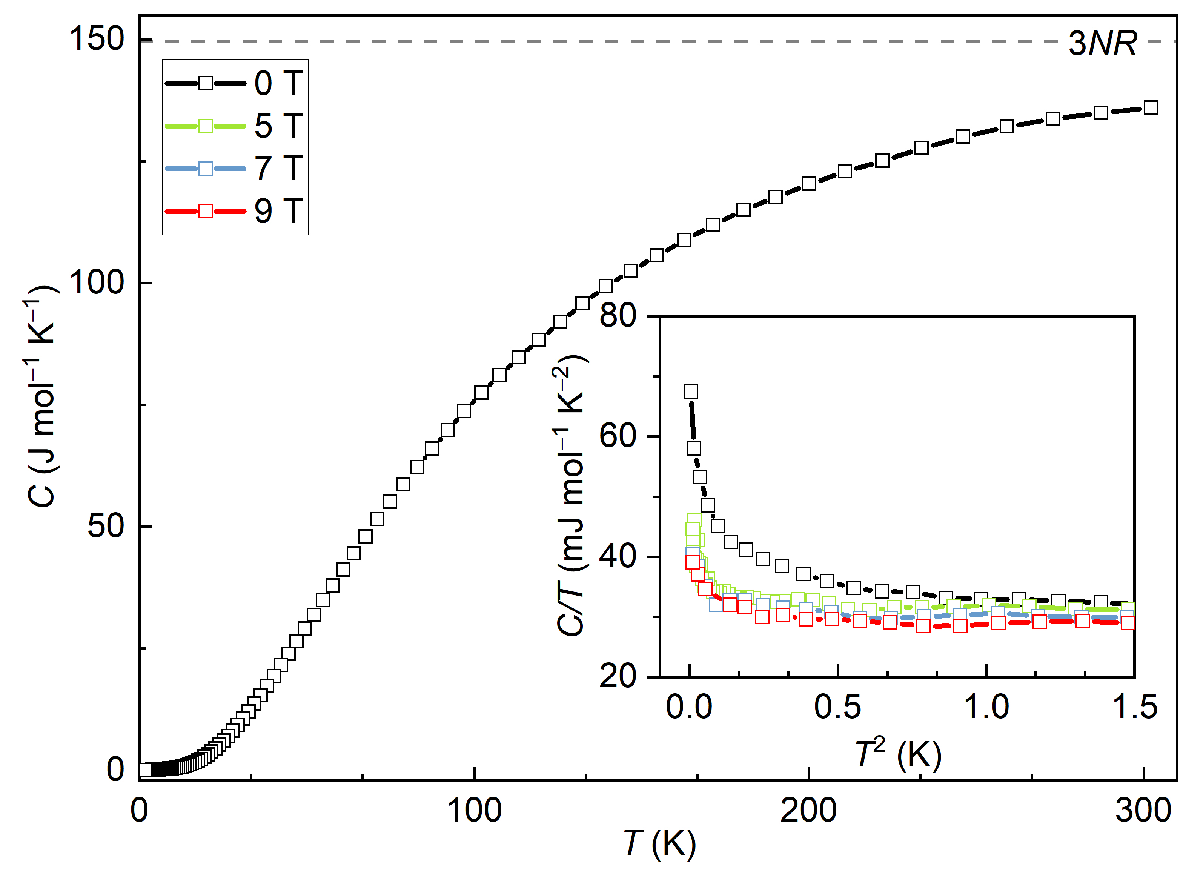}
    \centering
	\caption{Temperature dependence of the specific heat, $C(T)$, for the polycrystalline sample of ThCr$_2$Ge$_2$C. The inset plots $C/T$ versus $T^2$ under magnetic fields of 0 T, 5 T, 7 T, and 9 T.}
	\label{figS4_C}
\end{figure*}

\begin{figure*}[htbp]
	\includegraphics[width=11cm]{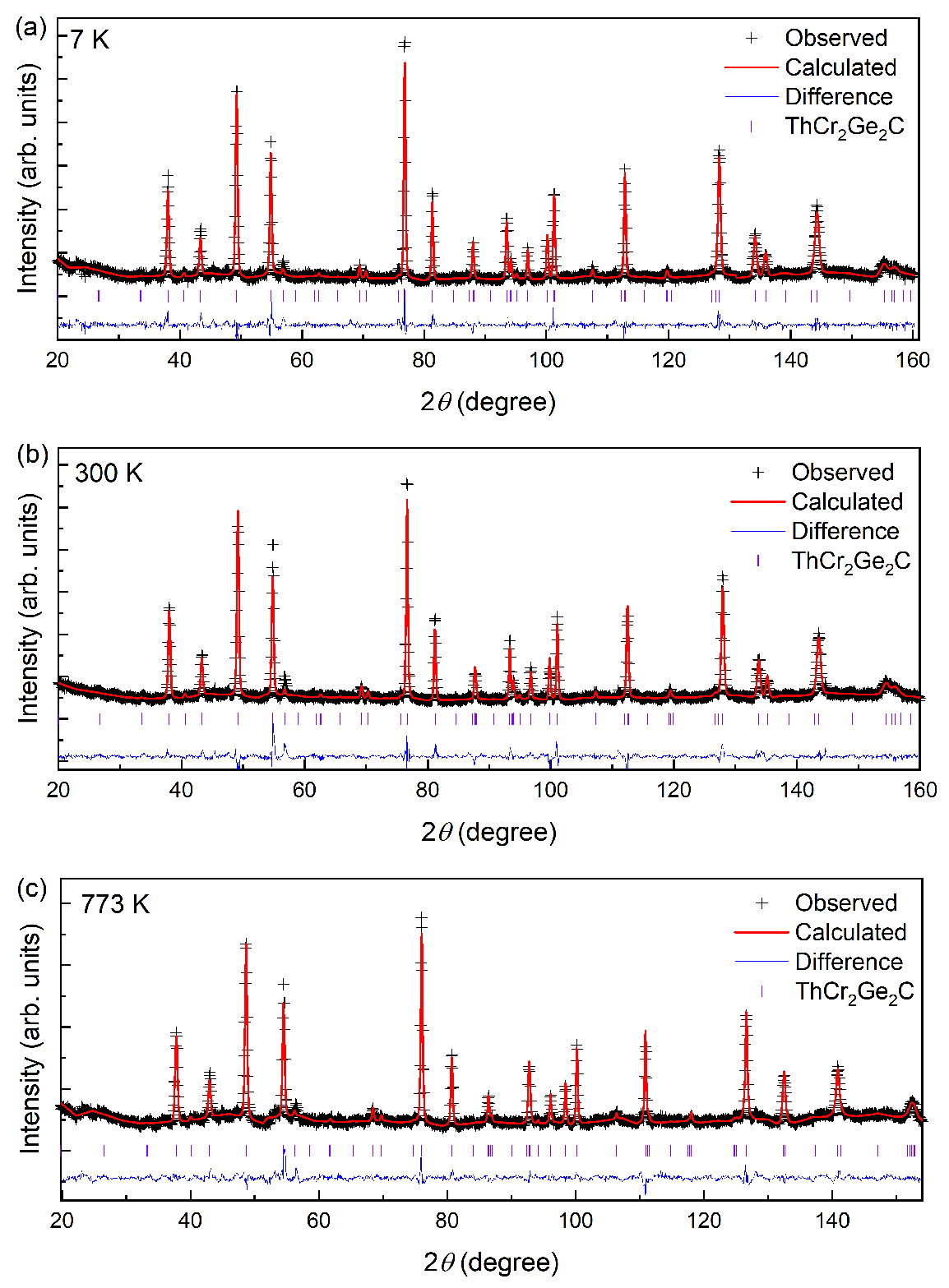}
    \centering
	\caption{The neutron powder diffraction patterns of ThCr$_2$Ge$_2$C at 7 K (a), 300 K (b), and 773 K (c).}
	\label{figS5_NPD}
\end{figure*}

\begin{figure*}[htbp]
	\includegraphics[width=13cm]{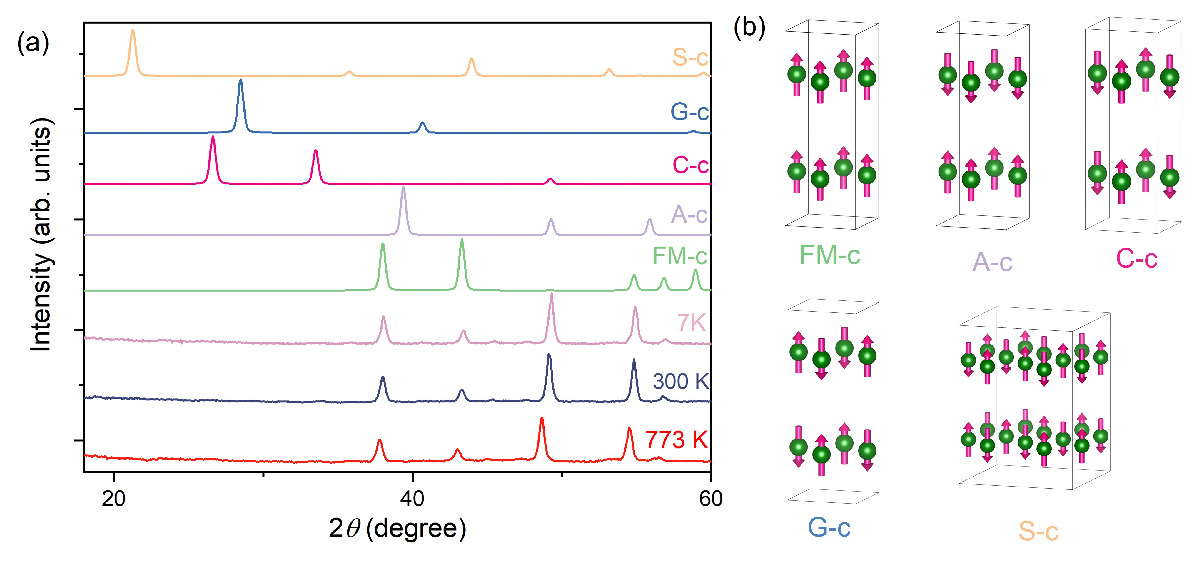}
    \centering
	\caption{(a) Simulated neutron diffraction patterns of different magnetic structures and experimental data for ThCr$_2$Ge$_2$C measured at 7~K, 300~K, and 773~K. (b) Diagrams of different types of magnetic structures. The spins of Cr are aligned the $c$-axis.}
	\label{figS6_compare}
\end{figure*}

\begin{figure*}[htbp]
	\includegraphics[width=11cm]{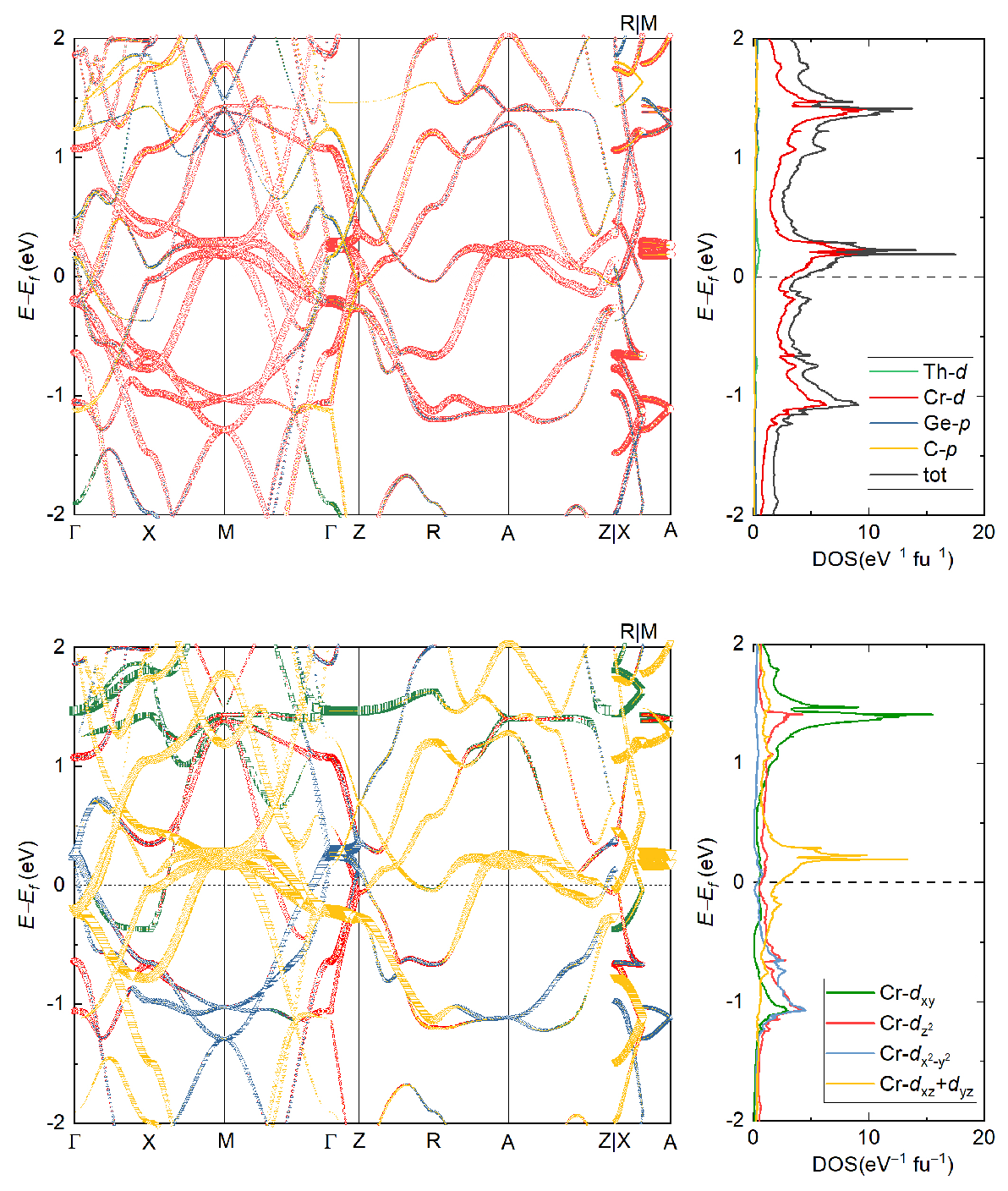}
    \centering
	\caption{(a) Orbital-projected band structure of ThCr$_2$Ge$_2$C along high-symmetry paths. (b) The corresponding total and orbital-projected density of states (DOS) at $U = 1$ eV. (c) Projected band structure of Cr-3$d$ orbitals in ThCr$_2$Ge$_2$C along high-symmetry paths. (d) The corresponding projected DOS of Cr-3$d$ orbitals at $U = 1$ eV.}
	\label{figS7_dosband}
\end{figure*}

\begin{figure*}[htbp]
	\includegraphics[width=10cm]{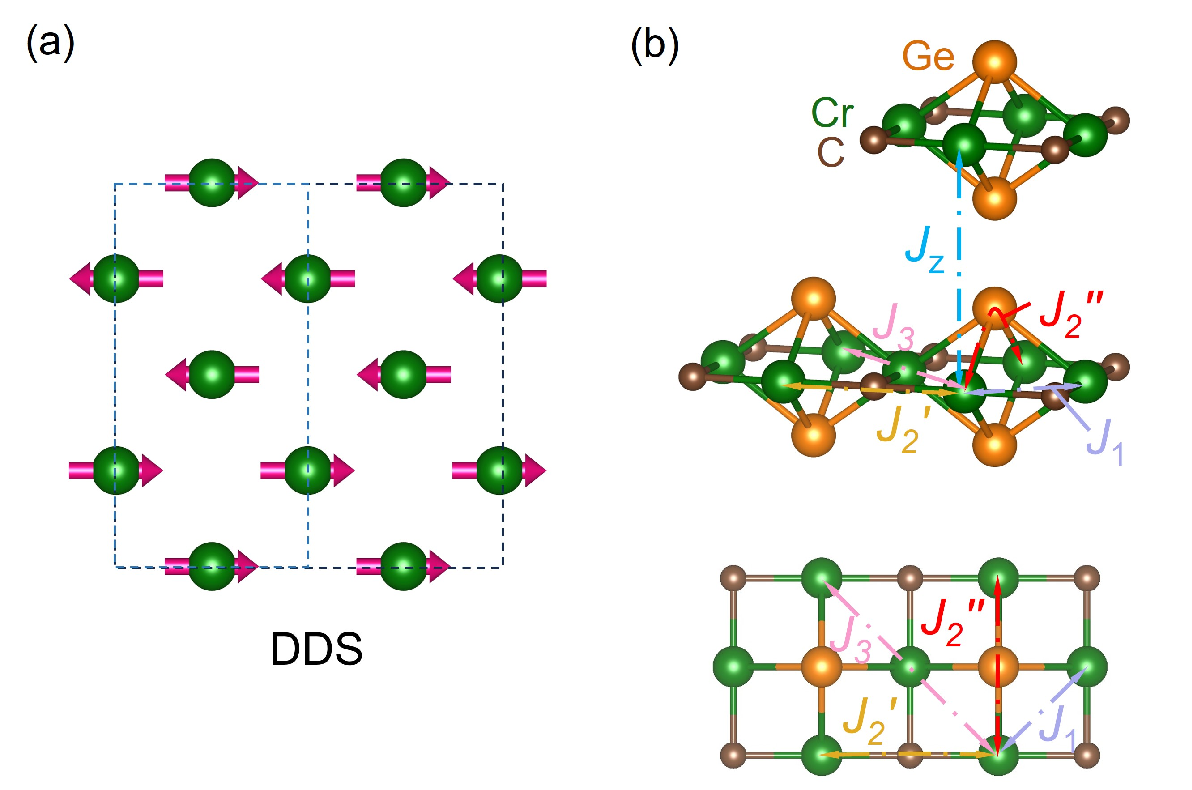}
    \centering
	\caption{(a) Diagram of the diagonal double stripe (DDS) magnetic structure with antiferromagnetic interlayer coupling, where the Cr spins are aligned within the $ab$-plane. The blue dashed-dotted square denotes the magnetic unit cell. (b) The magnetic exchange interactions between the Cr spins include an additional longer-range interaction, $J_3$.}
	\label{figS8_J3}
\end{figure*}

\begin{figure*}[htbp]
	\includegraphics[width=14cm]{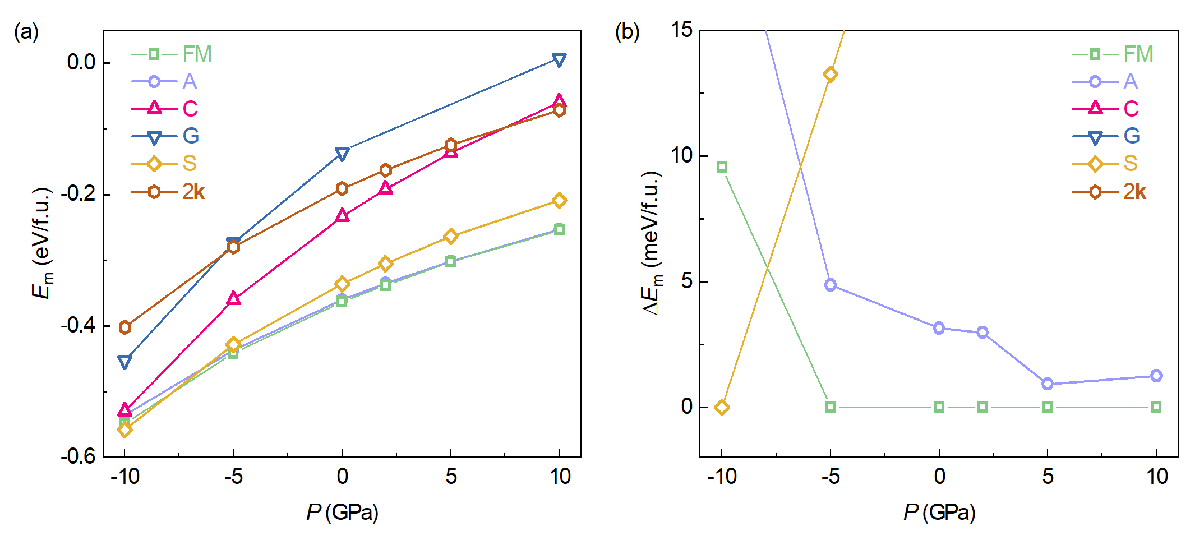}
    \centering
	\caption{(a) The calculated energy ($E_{\rm m}$) relative to the non-magnetic configuration for different magnetic configurations of ThCr$_2$Ge$_2$C under negative and positive pressures at $U = 1.5$~eV. (b) The calculated energy difference ($\Delta E_{\rm m}$) between the magnetic ground state and other magnetic structures at each pressure.}
	\label{figS9_dftHP}
\end{figure*}

\begin{table*}[htbp]
\caption{Crystallographic data of ThCr$_2$Ge$_2$C at 298~K from the single-crystal X-ray diffraction.}
\fontsize{10}{15}\selectfont
\begin{tabular}{p{1.3cm}<{\centering}p{1.3cm}<{\centering}p{1.3cm}<{\centering}p{1.3cm}<{\centering}p{1.3cm}<{\centering}p{1.3cm}<{\centering}p{1.3cm}<{\centering}
p{1.3cm}<{\centering}p{1.3cm}<{\centering}p{1.3cm}<{\centering}p{1.3cm}<{\centering}p{1.3cm}<{\centering}}
\hline \hline
\multicolumn{8}{l}{Empirical formula}                                                                  & \multicolumn{4}{l}{ThCr$_2$Ge$_2$C}                                                                                                                                                                \\
\multicolumn{8}{l}{Formula weight}                                                                     & \multicolumn{4}{l}{493.23~g/mol}                                                                                                                                                           \\
\multicolumn{8}{l}{Temperature}                                                                        & \multicolumn{4}{l}{298(2)~K}                                                                                                                                                               \\
\multicolumn{8}{l}{Wavelength}                                                                         & \multicolumn{4}{l}{0.71073~(\r{A})}                                                                                                                                                              \\
\multicolumn{8}{l}{Crystal system}                                                                     & \multicolumn{4}{l}{Tetragonal}                                                                                                                                                               \\
\multicolumn{8}{l}{Space group}                                                                        & \multicolumn{4}{l}{$P$4/$mmm$}                                                                                                                                                               \\
\multicolumn{8}{l}{\multirow{3}{*}{Unit cell dimensions}}                                              & \multicolumn{4}{l}{$a$ = 4.0973(3) (\r{A}), $\alpha$ = 90$^{\circ}$}                                                                                                                                               \\
\multicolumn{8}{l}{}                                                                                   & \multicolumn{4}{l}{$b$ = 4.0973(3) (\r{A}), $\beta$ = 90$^{\circ}$}                                                                                                                                                 \\
\multicolumn{8}{l}{}                                                                                   & \multicolumn{4}{l}{$c$ = 5.4369(8) (\r{A}), $\gamma$ = 90$^{\circ}$}                                                                                                                                                 \\
\multicolumn{8}{l}{Volume}                                                                             & \multicolumn{4}{l}{91.274(19) \r{A}$^3$}                                                                                                                                                          \\
\multicolumn{8}{l}{$Z$}                                                                                  & \multicolumn{4}{l}{1}                                                                                                                                                                        \\
\multicolumn{8}{l}{Density (calculated)}                                                               & \multicolumn{4}{l}{8.973~g/cm$^3$}                                                                                                                                                            \\
\multicolumn{8}{l}{Absorption coefficient}                                                             & \multicolumn{4}{l}{62.374~mm$^{-1}$}                                                                                                                                                            \\
\multicolumn{8}{l}{$F$(000)}                                                                             & \multicolumn{4}{l}{208}                                                                                                                                                                      \\
\multicolumn{8}{l}{Crystal size}                                                                       & \multicolumn{4}{l}{0.020 $\times$ 0.018 $\times$ 0.005 mm$^3$}                                                                                                                                              \\
\multicolumn{8}{l}{$\theta$ range for data collection}                                                        & \multicolumn{4}{l}{3.748$^{\circ}$ to 32.532$^{\circ}$}                                                                                                                                                       \\
\multicolumn{8}{l}{Index ranges}                                           & \multicolumn{4}{l}{$-$6 $\leq h \leq$ 6,   $-$6$\leq k \leq$6, $-$8$\leq l \leq$8}                                                                               \\
\multicolumn{8}{l}{Reflections collected}                                                              & \multicolumn{4}{l}{6000}                                                                                                                                                                     \\
\multicolumn{8}{l}{Independent reflections}                                                            & \multicolumn{4}{l}{132   {[}$R_{\rm int}$ = 0.0456{]}}                                                                                                                                                \\
\multicolumn{8}{l}{Completeness to $\theta$ = 25.242$^{\circ}$}                                                        & \multicolumn{4}{l}{100\%}                                                                                                                                                                    \\
\multicolumn{8}{l}{Refinement method}                                                                  & \multicolumn{4}{l}{Full-matrix   least-squares on $F^2$}                                                                                                                                        \\
\multicolumn{8}{l}{Data / restraints / parameters}                                                     & \multicolumn{4}{l}{132 /   0 / 11}                                                                                                                                                           \\
\multicolumn{8}{l}{Goodness-of-fit}                                                                    & \multicolumn{4}{l}{0.525}                                                                                                                                                                    \\
\multicolumn{8}{l}{Final $R^{\ast}$ indices [$I>3\sigma(I)$]}                                         & \multicolumn{4}{l}{$R_{\rm obs}$ = 0.0149, $wR_{\rm obs}$ = 0.0511}                                                                                                                                            \\
\multicolumn{8}{l}{$R$ indices [all data]}                                                           & \multicolumn{4}{l}{$R_{\rm all}$ = 0.0159, $wR_{\rm all}$ = 0.0531}                                                                                                                                            \\
\multicolumn{8}{l}{Extinction coefficient}                                                             & \multicolumn{4}{l}{NA}                                                                                                                                                                       \\
\multicolumn{8}{l}{Largest diff. peak and hole}                                                        & \multicolumn{4}{l}{1.345 and $-$3.477 e·\r{A}$^{-3}$}                                                                                                                                                 \\\hline
\multicolumn{12}{l}{$R = \sum \left| |F_o| - |F_c| \right| / \sum |F_o|, \, wR = \left( \sum w \left( |F_o|^2 - |F_c|^2 \right)^2 / \sum w \left( |F_o|^4 \right) \right)^{1/2},$ 
    and $w = 1 / \left( \sigma^2 (F_o^2) + (0.1000 F)^2 \right)$}                                                                                                                                       \\ \hline 
Atom                     & $x$                    & $y$                     & $z$                           & Occ.                  & $U_{\rm eq^{\ast}}$                   & $U_{\rm 11}$                    & $U_{\rm 22}$                    & $U_{\rm 33}$                   & $U_{\rm 12}$                  & $U_{\rm 13}$                 & $U_{\rm 23}$                 \\
Th                       & 0                     & 0                     & 0                           & 1                          & 5(1)                   & 5(1)                   & 5(1)                  & 6(1)                  & 0                   & 0                  & 0                  \\
Cr                       & 0.5                   & 0                     & 0.5                         & 1                          & 7(1)                   & 7(1)                   & 7(1)                  & 6(1)                  & 0                   & 0                  & 0                  \\
Ge                       & 0.5                   & 0.5                   & 0.7687(2)                   & 1                          & 9(1)                   & 18(1)                  & 4(1)                  & 6(1)                  & 0                   & 0                  & 0                  \\
C                        & 0                     & 0                     & 0.5                         & 1                          & 19(4)                  & 22(6)                  & 22(6)                 & 12(7)                 & 0                   & 0                  & 0                  \\\hline
\multicolumn{12}{l}{\makecell{$U_{\rm eq^{\ast}}$ is defined as one third of the trace of the orthogonalized $U_{ij}$ tensor. The anisotropic displacement factor \\exponent takes the form:
$-2\pi^2 \left[ h^2 a^{\ast 2} U_{\rm 11} + \dots + 2hk a^{\ast} b^{\ast} U_{\rm 12} \right]$. The unit of the displacement parameters is 0.001 \r{A}$^2$.}}

\\ \hline \hline
\end{tabular}
    \label{Crystal}
\end{table*}

\begin{table*}[htbp]
\caption{Refined structural of ThCr$_2$Ge$_2$C from X-ray diffraction and neutron powder diffraction measurements at different temperatures.}
\fontsize{10}{15}\selectfont
  \begin{tabular}{p{1.2cm}<{\centering}p{1.2cm}<{\centering}p{1.2cm}<{\centering}p{1.2cm}<{\centering}p{1.3cm}<{\centering}p{1.2cm}<{\centering}
  p{1.3cm}<{\centering}p{1.2cm}<{\centering}p{1.2cm}<{\centering}p{1.2cm}<{\centering}p{1.2cm}<{\centering}p{1.2cm}<{\centering}}
\hline \hline
\multicolumn{4}{c}{Compound}                     & \multicolumn{8}{c}{ThCr$_2$Ge$_2$C}                                                                                                                                                                                                                         \\
\multicolumn{4}{c}{Space group}                  & \multicolumn{8}{c}{$P$4/$mmm$ (No. 123)}                                                                                                                                                                                                                     \\  \hline
\multicolumn{4}{c}{Radiation}               & \multicolumn{2}{c|}{X-ray}                                           & \multicolumn{6}{c}{Neutron}                                                                                                                                                              \\
\multicolumn{4}{c}{Temperture}                   & \multicolumn{2}{c|}{300 K}                                         & \multicolumn{2}{c}{7 K}                                           & \multicolumn{2}{c}{300 K}                                         & \multicolumn{2}{c}{773 K}                    \\
\multicolumn{4}{c}{$a$ (\r{A})} & \multicolumn{2}{c|}{4.1018(1)}                                     & \multicolumn{2}{c}{4.0936(4)}                                     & \multicolumn{2}{c}{4.1008(4)}                                     & \multicolumn{2}{c}{4.1173(6)}                \\
\multicolumn{4}{c}{$c$ (\r{A})} & \multicolumn{2}{c|}{5.4471(1)}                                     & \multicolumn{2}{c}{5.4317(6)}                                     & \multicolumn{2}{c}{5.4480(6)}                                     & \multicolumn{2}{c}{5.5055(1)}                \\
\multicolumn{4}{c}{$R_{\mathrm{wp}}$ ($\%$)}     & \multicolumn{2}{c|}{8.05}                                          & \multicolumn{2}{c}{7.88}                                          & \multicolumn{2}{c}{7.08}                                          & \multicolumn{2}{c}{6.32}                     \\
\multicolumn{4}{c}{$R_{\mathrm{e}}$ ($\%$)}      & \multicolumn{2}{c|}{3.42}                                          & \multicolumn{2}{c}{4.06}                                          & \multicolumn{2}{c}{3.90}                                          & \multicolumn{2}{c}{4.17}                     \\
\multicolumn{4}{c}{$S$}                   & \multicolumn{2}{c|}{2.35}                                          & \multicolumn{2}{c}{1.94}                                          & \multicolumn{2}{c}{1.82}                                          & \multicolumn{2}{c}{1.52}                     \\ \hline \hline
Atom  & site  & $x$  & \multicolumn{1}{c|}{$y$}  & $z$       & \multicolumn{1}{c|}{$B$ ({\AA}$^2$)} & $z$       & \multicolumn{1}{c|}{$B$ ({\AA}$^2$)} & $z$       & \multicolumn{1}{c|}{$B$ ({\AA}$^2$)} & $z$       & {$B$ ({\AA}$^2$)} \\ \hline
Th    & 1$a$  & 0    & \multicolumn{1}{c|}{0}    & 0         & \multicolumn{1}{c|}{0.02(1)}                              & 0         & \multicolumn{1}{c|}{0.3(1)}                              & 0         & \multicolumn{1}{c|}{0.5(1)}                              & 0         & 0.3(1)                              \\
Cr    & 2$e$  & 0.5    & \multicolumn{1}{c|}{0}    & 0.5       & \multicolumn{1}{c|}{0.31(7)}                            & 0.5         & \multicolumn{1}{c|}{0.4(1)}                            & 0.5         & \multicolumn{1}{c|}{0.8(1)}                              & 0.5         & 1.0(1)                              \\
Ge    & 2$h$  & 0.5  & \multicolumn{1}{c|}{0.5}  & 0.7724(4) & \multicolumn{1}{c|}{0.49(6)}                              & 0.7717(7) & \multicolumn{1}{c|}{0.04(9)}                             & 0.7712(7) & \multicolumn{1}{c|}{0.4(1)}                              & 0.7738(7) & 0.3(1)                              \\
C     & 1$b$  & 0    & \multicolumn{1}{c|}{0}    & 0.5       & \multicolumn{1}{c|}{3.13(2)}                              & 0.5       & \multicolumn{1}{c|}{0.7(1)}                              & 0.5       & \multicolumn{1}{c|}{1.1(1)}                              & 0.5       & 1.5(2)                              \\ \hline \hline
\end{tabular}
    \label{NPD}
\end{table*}

\begin{table*}[htbp]
\centering
\fontsize{7.5}{16}\selectfont
\caption{The calculated magnetic energy $E_\mathrm{m}$ (in meV/f.u.) relative to the non-magnetic (NM) state and the corresponding magnetic moment $\mu_{\mathrm{Cr}}$ (in $\mu_\mathrm{B}$) for different magnetic structures of ThCr$_2$Ge$_2$C. The magnetic ground state and its magnetic moment for each $U$ value are emphasized in bold.}
\begin{tabular}{p{3cm}<{\centering}|p{1.5cm}<{\centering}p{1.5cm}<{\centering}p{1.5cm}<{\centering}p{1.5cm}<{\centering}
p{1.5cm}<{\centering}p{1.5cm}<{\centering}p{1.5cm}<{\centering}p{1.5cm}<{\centering}p{1.5cm}<{\centering}
p{1.5cm}<{\centering}p{1.5cm}<{\centering}p{1.5cm}<{\centering}p{1.5cm}<{\centering}p{1.5cm}<{\centering}
p{1.5cm}<{\centering}p{1.5cm}<{\centering}p{1.5cm}<{\centering}p{1.5cm}<{\centering}p{1.5cm}<{\centering}
p{1.5cm}<{\centering}p{1.5cm}<{\centering}p{1.5cm}<{\centering}p{1.5cm}<{\centering}p{1.cm}<{\centering}
p{1.5cm}<{\centering}p{1.5cm}<{\centering}}
\hline \hline
\multirow{3}{*}{Magnetic structure} & \multicolumn{14}{c}{$U$ (eV)}                                                                                                                                                                                                                                                                                                                                                                                                                                                                                                                                   \\ \cline{2-15} 
                                    & \multicolumn{2}{c|}{0.0}                                                       & \multicolumn{2}{c|}{0.5}                                                       & \multicolumn{2}{c|}{1.0}                                                       & \multicolumn{2}{c|}{1.5}                                                       & \multicolumn{2}{c|}{2.0}                                                       & \multicolumn{2}{c|}{2.5}                                                       & \multicolumn{2}{c}{3.0}                                   \\ \cline{2-15} 
                                    & \multicolumn{1}{c|}{$E_\mathrm{m}$} & \multicolumn{1}{c|}{$\mu_{\mathrm{Cr}}$} & \multicolumn{1}{c|}{$E_\mathrm{m}$} & \multicolumn{1}{c|}{$\mu_{\mathrm{Cr}}$} & \multicolumn{1}{c|}{$E_\mathrm{m}$} & \multicolumn{1}{c|}{$\mu_{\mathrm{Cr}}$} & \multicolumn{1}{c|}{$E_\mathrm{m}$} & \multicolumn{1}{c|}{$\mu_{\mathrm{Cr}}$} & \multicolumn{1}{c|}{$E_\mathrm{m}$} & \multicolumn{1}{c|}{$\mu_{\mathrm{Cr}}$} & \multicolumn{1}{c|}{$E_\mathrm{m}$} & \multicolumn{1}{c|}{$\mu_{\mathrm{Cr}}$} & \multicolumn{1}{c|}{$E_\mathrm{m}$} & $\mu_{\mathrm{Cr}}$ \\ \hline
FM                                  & \multicolumn{1}{c|}{$-$53.80}         & \multicolumn{1}{c|}{1.05}                & \multicolumn{1}{c|}{$-$132.06}        & \multicolumn{1}{c|}{1.22}                & \multicolumn{1}{c|}{$-$232.99}        & \multicolumn{1}{c|}{1.40}                & \multicolumn{1}{c|}{\textbf{$-$363.19}}        & \multicolumn{1}{c|}{\textbf{1.63}}                & \multicolumn{1}{c|}{\textbf{$-$530.78}}        & \multicolumn{1}{c|}{\textbf{1.90}}                & \multicolumn{1}{c|}{$-$743.61}        & \multicolumn{1}{c|}{2.19}                & \multicolumn{1}{c|}{$-$1030.89}       & 2.63                \\ \hline
A                                   & \multicolumn{1}{c|}{\textbf{$-$64.95}}         & \multicolumn{1}{c|}{\textbf{0.95}}                & \multicolumn{1}{c|}{\textbf{$-$137.24}}        & \multicolumn{1}{c|}{\textbf{1.19}}                & \multicolumn{1}{c|}{\textbf{$-$236.85}}        & \multicolumn{1}{c|}{\textbf{1.37}}                & \multicolumn{1}{c|}{$-$360.05}        & \multicolumn{1}{c|}{1.56}                & \multicolumn{1}{c|}{$-$513.92}        & \multicolumn{1}{c|}{1.74}                & \multicolumn{1}{c|}{$-$697.48}        & \multicolumn{1}{c|}{1.98}                & \multicolumn{1}{c|}{$-$926.16}        & 2.25                \\ \hline
C                                   & \multicolumn{2}{c|}{Converge to FM}                                            & \multicolumn{2}{c|}{Converge to FM}                                            & \multicolumn{2}{c|}{Converge to FM}                                            & \multicolumn{1}{c|}{$-$233.25}        & \multicolumn{1}{c|}{2.02}                & \multicolumn{1}{c|}{$-$458.19}        & \multicolumn{1}{c|}{2.28}                & \multicolumn{1}{c|}{$-$727.93}        & \multicolumn{1}{c|}{2.50}                & \multicolumn{1}{c|}{$-$1032.32}       & 2.66                \\ \hline
G                                   & \multicolumn{2}{c|}{Converge to A}                                             & \multicolumn{2}{c|}{Converge to A}                                             & \multicolumn{2}{c|}{Converge to A}                                             & \multicolumn{1}{c|}{$-$136.10}        & \multicolumn{1}{c|}{2.02}                & \multicolumn{1}{c|}{$-$372.57}        & \multicolumn{1}{c|}{2.31}                & \multicolumn{1}{c|}{$-$652.27}        & \multicolumn{1}{c|}{2.51}                & \multicolumn{1}{c|}{$-$965.34}        & 2.66                \\ \hline
S                                   & \multicolumn{1}{c|}{$-$38.26}         & \multicolumn{1}{c|}{0.74}                & \multicolumn{1}{c|}{$-$98.20}         & \multicolumn{1}{c|}{1.02}                & \multicolumn{1}{c|}{$-$188.64}        & \multicolumn{1}{c|}{1.52}                & \multicolumn{1}{c|}{$-$336.03}        & \multicolumn{1}{c|}{1.83}                & \multicolumn{1}{c|}{$-$525.23}        & \multicolumn{1}{c|}{2.07}                & \multicolumn{1}{c|}{\textbf{$-$753.97}}        & \multicolumn{1}{c|}{\textbf{2.32}}                & \multicolumn{1}{c|}{\textbf{$-$1034.34}}       & \textbf{2.77}               \\ \hline
2\textbf{k}                        & \multicolumn{2}{c|}{Converge to NM}                                            & \multicolumn{2}{c|}{Converge to NM}                                            & \multicolumn{1}{c|}{$-$62.71}         & \multicolumn{1}{c|}{1.42}                & \multicolumn{1}{c|}{$-$190.57}        & \multicolumn{1}{c|}{1.66}                & \multicolumn{1}{c|}{$-$350.29}        & \multicolumn{1}{c|}{1.92}                & \multicolumn{1}{c|}{$-$553.32}        & \multicolumn{1}{c|}{2.28}                & \multicolumn{1}{c|}{$-$826.26}        & 2.71                \\ \hline \hline
\end{tabular}
    \label{dft}
\end{table*}

\begin{table}[t]
\centering
\fontsize{9}{15}\selectfont
\caption{The calculated spin exchange interactions and the $\alpha$ ($= (J'_2 + J''_2) / 2J_1 $) values in ThCr$_2$Ge$_2$C for $U = 1.5$ and 2~eV, when the $J_3$ term is included.}
   \begin{tabular}{c|ccccc|c}
   \hline \hline
\multirow{2}{*}{$U$ (eV)} & \multicolumn{5}{c|}{$J_i$ (meV/$S^2$)}                                                                                                & \multirow{2}{*}{$\alpha$} \\ \cline{2-6}
                          & \multicolumn{1}{c|}{$J_1$}     & \multicolumn{1}{c|}{$J'_2$} & \multicolumn{1}{c|}{$J''_2$}  & \multicolumn{1}{c|}{$J_3$}   & $J_z$   &                           \\ \hline
1.5                       & \multicolumn{1}{c|}{$-$27.99} & \multicolumn{1}{c|}{33.05}  & \multicolumn{1}{c|}{$-$11.07} & \multicolumn{1}{c|}{$-$1.66} & $-$0.79 & $-$0.39                   \\ \hline
2                         & \multicolumn{1}{c|}{$-$17.67}  & \multicolumn{1}{c|}{45.00}  & \multicolumn{1}{c|}{$-$24.51} & \multicolumn{1}{c|}{0.63}    & $-$4.22 & $-$0.58                   \\ \hline\hline
\end{tabular}
    \label{J3}
\end{table}

\begin{table*}[htbp]
\centering
\fontsize{7.5}{13}\selectfont
\caption{The calculated magnetic energy $E_\mathrm{m}$ (in meV/f.u.) relative to the non-magnetic (NM) state and the corresponding magnetic moment $\mu_{\mathrm{Cr}}$ (in $\mu_\mathrm{B}$) for different magnetic structures of ThCr$_2$Ge$_2$C under positive and negative pressures at $U =$~1.5~eV. The magnetic ground state and its magnetic moment for each pressure are emphasized in bold. The last row $\alpha$ is defined by $\alpha = (J'_2 + J''_2) / 2J_1 $.}
\begin{tabular}{p{3cm}<{\centering}|p{1.5cm}<{\centering}p{1.5cm}<{\centering}p{1.5cm}<{\centering}p{1.5cm}<{\centering}
p{1.5cm}<{\centering}p{1.5cm}<{\centering}p{1.5cm}<{\centering}p{1.5cm}<{\centering}p{1.5cm}<{\centering}
p{1.5cm}<{\centering}}
\hline \hline
\multirow{3}{*}{Magnetic structure} & \multicolumn{10}{c}{$P$$~$(GPa)}                                                                                                                                                                                                                                                                                                              \\ \cline{2-11} 
                                    & \multicolumn{2}{c|}{$-$10.00}                                 & \multicolumn{2}{c|}{$-$5.00}                                  & \multicolumn{2}{c|}{2.00}                                                         & \multicolumn{2}{c|}{5.00}                                                         & \multicolumn{2}{c}{10.00}                \\ \cline{2-11} 
                                    & $E_\mathrm{m}$     & \multicolumn{1}{c|}{$\mu_{\mathrm{Cr}}$} & $E_\mathrm{m}$     & \multicolumn{1}{c|}{$\mu_{\mathrm{Cr}}$} & \multicolumn{1}{l}{$E_\mathrm{m}$}     & \multicolumn{1}{c|}{$\mu_{\mathrm{Cr}}$} & \multicolumn{1}{l}{$E_\mathrm{m}$}     & \multicolumn{1}{c|}{$\mu_{\mathrm{Cr}}$} & $E_\mathrm{m}$     & $\mu_{\mathrm{Cr}}$ \\ \hline
FM                                  & $-$548.27          & \multicolumn{1}{c|}{1.95}                & \textbf{$-$441.33} & \multicolumn{1}{c|}{\textbf{1.81}}       & \multicolumn{1}{l}{\textbf{$-$337.98}} & \multicolumn{1}{c|}{\textbf{1.58}}       & \multicolumn{1}{l}{\textbf{$-$302.43}} & \multicolumn{1}{c|}{\textbf{1.52}}       & \textbf{$-$254.09} & \textbf{1.41}       \\
A                                   & $-$536.01          & \multicolumn{1}{c|}{1.81}                & $-$436.47          & \multicolumn{1}{c|}{1.66}                & \multicolumn{1}{l}{$-$335.00}          & \multicolumn{1}{c|}{1.51}                & \multicolumn{1}{l}{$-$301.51}          & \multicolumn{1}{c|}{1.45}                & $-$252.84          & 1.38                \\
C                                   & $-$529.77          & \multicolumn{1}{c|}{2.37}                & $-$360.29          & \multicolumn{1}{c|}{2.20}                & \multicolumn{1}{l}{$-$192.09}          & \multicolumn{1}{c|}{1.97}                & \multicolumn{1}{l}{$-$136.54}          & \multicolumn{1}{c|}{1.88}                & $-$59.58           & 1.72                \\
G                                   & $-$452.79          & \multicolumn{1}{c|}{2.37}                & $-$274.04          & \multicolumn{1}{c|}{2.22}                & \multicolumn{2}{c|}{Converge to A}                                                & \multicolumn{2}{c|}{Converge to A}                                                & 7.38               & 1.06                \\
S                                   & \textbf{$-$557.82} & \multicolumn{1}{c|}{\textbf{2.17}}       & $-$428.07          & \multicolumn{1}{c|}{1.98}                & \multicolumn{1}{l}{$-$305.44}          & \multicolumn{1}{c|}{1.78}                & \multicolumn{1}{l}{$-$263.32}          & \multicolumn{1}{c|}{1.69}                & $-$208.77          & 1.52                \\
\textbf{2k}                         & $-$402.06          & \multicolumn{1}{c|}{2.09}                & $-$279.60          & \multicolumn{1}{c|}{1.84}                & \multicolumn{1}{l}{$-$162.89}          & \multicolumn{1}{c|}{1.60}                & \multicolumn{1}{l}{$-$124.76}          & \multicolumn{1}{c|}{1.53}                & $-$71.58           & 1.41                \\ \hline\hline
$J_1$                               & \multicolumn{2}{c|}{$-$2.31}                                  & \multicolumn{2}{c|}{$-$10.13}                                 & \multicolumn{2}{c|}{$-$18.24}                                                     & \multicolumn{2}{c|}{$-$20.74}                                                     & \multicolumn{2}{c}{$-$24.31}             \\
$J'_2$                              & \multicolumn{2}{c|}{37.14}                                    & \multicolumn{2}{c|}{37.68}                                    & \multicolumn{2}{c|}{34.82}                                                        & \multicolumn{2}{c|}{34.77}                                                        & \multicolumn{2}{c}{12.98}                \\
$J''_2$                             & \multicolumn{2}{c|}{$-$31.18}                                 & \multicolumn{2}{c|}{$-$29.09}                                 & \multicolumn{2}{c|}{$-$24.79}                                                     & \multicolumn{2}{c|}{$-$23.45}                                                     & \multicolumn{2}{c}{$-$21.31}             \\
$J_z$                               & \multicolumn{2}{c|}{$-$3.07}                                  & \multicolumn{2}{c|}{$-$3.07}                                  & \multicolumn{2}{c|}{$-$0.75}                                                      & \multicolumn{2}{c|}{$-$0.23}                                                      & \multicolumn{2}{c}{$-$0.32}              \\ \hline\hline
$\alpha$                            & \multicolumn{2}{c|}{$-$1.29}                                  & \multicolumn{2}{c|}{$-$0.42}                                  & \multicolumn{2}{c|}{$-$0.28}                                                      & \multicolumn{2}{c|}{$-$0.27}                                                      & \multicolumn{2}{c}{0.17}                 \\ \hline\hline
\end{tabular}
    \label{dftHP}
\end{table*}
\end{widetext}
\FloatBarrier

\bibliography{ThCr2Ge2C}
\end{document}